\documentclass[11pt,mnsc, nonblindrev]{informs3_noMNSC_removethis_}
\usepackage[T1]{fontenc}
\usepackage[latin9]{inputenc}
\usepackage{array}
\usepackage{url}
\usepackage{multirow}
\usepackage{amsmath}
\usepackage{amssymb}
\usepackage{graphicx}
\usepackage[authoryear]{natbib}
\usepackage[unicode=true,pdfusetitle,
 bookmarks=true,bookmarksnumbered=true,bookmarksopen=false,
 breaklinks=false,pdfborder={0 0 0},pdfborderstyle={},backref=false,colorlinks=false]
 {hyperref}

\makeatletter

\providecommand{\tabularnewline}{\\}

\OneAndAHalfSpacedXI

\usepackage{natbib}
 \bibpunct[, ]{(}{)}{,}{a}{}{,}%
 \def\bibfont{\small}%
\TheoremsNumberedThrough     
\ECRepeatTheorems

\EquationsNumberedThrough    

\makeatother

\begin{document}
\RUNTITLE{Musalem, Shang and Song} 

\TITLE{An Empirical Study of Customer Spillover Learning about Service Quality}

\ARTICLEAUTHORS{		
    \AUTHOR{Andr\'{e}s Musalem}
	\AFF{Industrial Engineering Department, University of Chile; \EMAIL{amusalem@dii.uchile.cl}}
    \AUTHOR{Yan Shang}
	\AFF{Facebook, Inc.; \EMAIL{yanshang@fb.com}}
	\AUTHOR{Jing-Sheng Song}
	\AFF{Fuqua School of Business, Duke University; \EMAIL{jingsheng.song@duke.edu}}}

\ABSTRACT{"Spillover" learning  is defined as  customers' learning about the quality of a service (or product) from their previous experiences with similar yet not identical services. In this paper, we propose a novel, parsimonious and general Bayesian hierarchical learning framework for estimating customers' spillover learning. We apply our model to a one-year shipping/sales historical data provided by a world-leading third party logistics company and study how customers' experiences from shipping on a particular route affect their future  decisions about shipping not only on that route, but also on other routes serviced by the same logistics company. Our empirical results are consistent with information spillovers driving customer choices. Customers also display an asymmetric response such that they are more sensitive to delays than early deliveries. In addition, we find that customers are risk averse being more sensitive to their uncertainty about the mean service quality than to the intrinsic variability of the service. Finally, we develop policy simulation studies to show the importance of accounting for customer learning when a firm considers service quality improvement decisions. }


\HISTORY{This paper was first submitted on July 18, 2016.}

\maketitle

\section{Introduction }

A central goal of service firms is to offer high levels of customer
satisfaction while maintaining a low cost. To contribute to the achievement
of this goal, the service operations management literature has focused
on developing normative models to minimize operational (such as staffing
and inventory) costs to attain a given service level. However, there
is relatively scant research on how to choose an appropriate target
service level. Indeed, the choice of service levels can be seen as
a balancing act: higher service levels improve customer satisfaction
and demand, but they are also likely to increase service costs and
therefore negatively impact demand. Determining the effect of service
levels on customer utility and service costs requires measuring the
value that customers assign to an objective service level improvement
and how this additional value translates into incremental revenues
and profits. In this regard, empirical research focused on how customer
purchasing behavior is affected by a firm's service levels is particularly
valuable. Accordingly, the focus of this paper is to develop a structural
model to enable such measurement in a context where customers learn
about the quality of service through repeated encounters with a service
provider in a business-to-business (B2B) setting. 

Our modeling framework is motivated by and built on a one-year international
cargo shipping and sales data from individual customers of a world-leading
third-party logistics company. The dataset contains information about
the quality of the service (e.g., delivery times and lengths of delays)
received by customers each time they interacted with this provider.
Air transport delivers goods that are time-sensitive, expensive, perishable
or used in just-in-time supply networks, at competitive prices to
customers worldwide. As such, delivery time and reliability is a critical
measure of service quality. According to a 2010 report of Infosys,
\textquotedblleft carrier delays and non-performance on delivery\textquotedblright{}
is ranked as the leading risk in the logistics industry. In a 2014
survey conducted by the International Air Transport Association (IATA)
to major freight forwarders and their customers, low reliability is
perceived as the second most important factor (next to transportation
cost) driving the transport model shift from air to sea. 

Modeling the value of service quality to customers introduces several
challenges. By leveraging our dataset, we are able to overcome each
of the ones described next. First, the quality of service can only
be fully observed by the customer when the service is finished. For
instance, in our air cargo shipping context, when choosing a service
provider customers do not know whether their cargo will be delivered
to the destination on time because the actual delivery performance
can only be assessed after the cargo has been delivered. This uncertainty
makes the standard random utility framework unsuitable because it
typically assumes that customers know the attributes of all alternatives
perfectly before making a choice, such as in \citet{aksin_structural_2013}.
In contrast, a learning model is more suitable to our context (e.g.,
\citealt{ching_invited_2013}), because they allow customers to have
incomplete information about product attributes. Specifically, customers
form expectations about these attributes from one or several information
sources such as past experiences with the product, word of mouth and
advertisements, and then use their beliefs about service performance
in their purchase decisions. In this paper, we follow this approach
and study customers' learning process about air cargo shipping service
quality (i.e., on-time delivery). Our data set has an advantage over
most quality learning models in the empirical literature, where service
quality for each experience is not observed and hence is treated as
a latent construct (e.g., \citealt{erdem_empirical_1998}). 

Second, most studies using learning models assume that the information
about a certain product comes exclusively from experiences with the
same product. Such assumption works well when a consumer has had multiple
and frequent experiences with each product. When this is not the case,
consumers might learn from multiple experiences with similar, but
not necessarily identical products. We refer to this belief updating
mechanism as \textit{``spillover''} learning. In our application,
assume for example that a customer has ordered cargo shipping services
several times from a logistics company on a route from airport $A$
to airport $B$ and that the flights have often been either delayed
at the departure or the cargo has failed to be loaded onto the flight.\footnote{In this study, a route is defined as a directed pair from origin airport
$A$ to destination airport $B$, $A\to B$, and it is different from
$B\to A$. } Then, when the customer considers sending a new shipment through
a new route (e.g., from airport $C$ to airport $D$), the customer
may also expect some delays because of the reliability experienced
when using the same logistics company in the past. These information
spillovers could be further enhanced if the two routes share certain
characteristics. For example, the following two paths, $A\rightarrow B$
and $A\rightarrow C$, share the same origin airport and this would
make the experience on route $A\rightarrow B$ potentially more informative
about the delays to be anticipated when shipping on route $A\rightarrow D$,
than when anticipating the reliability of shipments on route $C\rightarrow D$. 

In the literature, a similar phenomenon has been studied (e.g., \citet{erdem_empirical_1998,sridhar_investigating_2012}).
For example, \citet{erdem_empirical_1998} considers a model where
the prior quality beliefs are correlated across products in different
categories from the same ``umbrella brand'' (i.e., a brand that
operates in multiple categories) through the use of a covariance matrix,
and finds evidence that consumers learn from experiences across umbrella
brands in the toothpaste and toothbrush categories. In this paper,
we explore a new methodology that relies on a Bayesian hierarchical
model, which yields a more parsimonious formulation and can be easily
extended to accommodate a variety of learning spillover processes.
This is also possible because we have access to quality information
and hence we don't need to impute this information as in the case
in most papers in the quality learning literature.

Third, data on service quality is often subjective rather than objective.
It is thus not surprising that most of the extant studies about customer
sensitivity to service quality rely on survey data (e.g., \citealt{bolton_dynamic_1998}).
While surveys are useful in providing information about how customers
perceive the quality of the service, it is not clear how to link subjective
service quality assessments to an operational decision or service
level. The delivery times and on-time frequency in our dataset, however,
correspond to objective service quality metrics, making our model
more applicable for operational decision making. 

In terms of our results, applying our methodology to model customer
behavior in the international air cargo transport industry, we find
evidence consistent with information spillovers driving future purchase
decisions. This result is important when estimating the overall impact
of a service improvement or failure, because a change in one particular
service may also affect other related services. Thus, our model can
be used to identify the overall effect of an operational change in
all products/services accounting for learning spillovers. We also
find that service quality beliefs have an economically significant
impact on the purchasing decisions of shipping services from the logistics
company. Moreover, this effect is asymmetric \textemdash{} customers
are approximately 1.7 times more sensitive to ``poor'' than to ``good''
service experiences. Similar to the results in the previous literature,
we find customers to be risk averse. In particular, we find that customers
are not only averse to experience variability, but also to belief
uncertainty (i.e., customer's uncertainty about their beliefs, see
$\mathsection$4.3 for definition and details). In fact, we find that
belief uncertainty affects customers' utilities more compared to experience
variability. To the best of our knowledge, we are the first to separately
estimate the effect of these two types of service quality uncertainty
on customer behavior. We show that these results have important implications
for designing service improvement strategies, which should be based
on a good understanding of the learning rules that customers use and
a careful balance of the direct and indirect (i.e., spillover) effect
of changes in service quality against the corresponding operational
costs. 

The rest of the paper is organized as follows. We position our work
within the literature in Section \ref{sec:lit}. In Section \ref{sec:data}
we describe the empirical setting. Section \ref{sec:model} presents
the learning and demand model and discusses the identification of
the model parameters. In Section \ref{sec:Results-and-Discussion}
we report the results. Section \ref{sec: Applications} discusses
managerial implications of our model and its empirical results through
counterfactual experiments. Section \ref{sec: Conclusions} concludes
the paper and presents future research directions. 

\section{Contributions to the Literature\label{sec:lit}}

Our study contributes to three steams of research. In this section
we provide a brief review of the literature that is most closely related
to our study in each of these streams. 

\subsection{Contributions to the Empirical Service Operations Management Literature}

Recent empirical studies have explored the problem of estimating the
effect of service quality on demand. For example, \citet{brown_statistical_2005}
and \citet{aksin_structural_2013} study customers' abandonment behavior
in the context of call centers, whereas \citet{batt_waiting_2015}
study patients abandonment behavior in the context of hospital emergency
room. \citet{allon_how_2011} quantify the effect of waiting time
on sales, using market level data from the fast food industry and
\citet{lu_measuring_2013} study the effect of waiting time and queue
length on customers' purchase decisions in a supermarket deli section.
Other studies have focused on different service quality measures including
delivery time in quick service restaurants \citep{cho_value_2015},
signal quality in video streaming services \citep{sridhar_investigating_2012},
on-time performance of airlines \citep{grewal_customer_2010} and
order fill-rate in a supplier-retailer supply chain \citep{craig_impact_2014}. 

Our paper contributes to this stream of research by explicitly modeling
the customers' learning process about service quality, such that their
beliefs about the quality of future experiences may depend on past
experiences. The previous empirical work in service operations management
typically uses a static approach to consider customers' beliefs about
service quality. For example, \citet{lu_measuring_2013} assume that
customers form beliefs about service quality when they arrive at the
physical queue based on the current observed length of the queue;
\citet{aksin_structural_2013} assume the customers (i.e. callers)
correctly anticipate service quality (i.e., the probability of receiving
service in a certain period) based on their past experiences and that
this service quality is common knowledge among all customers; \citet{batt_waiting_2015}
considers a hospital emergency department and argues that it is not
clear what customers (i.e. patients) can learn from what they observe
while waiting, so they investigate how the factors that customers
observe and experience in the waiting room impacts their abandonment
behavior. Instead of relying on this static assumption about consumer
beliefs, we explicitly allow consumer expectations to evolve over
time as each customer gains additional experiences with the service
provider. Allowing customers to form beliefs based only on what they
observe immediately before choosing a service provider can be a reasonable
assumption when the service environment is observable (e.g., when
the customer observes the length of the queue as in \citet{lu_measuring_2013}).
However, in other contexts this is not the case, such as in call centers
where a caller does not directly observe the state of the queue, and
in air cargo transportation where a shipper chooses a service provider
before observing the congestion level that his shipment will face
once the cargo is picked up. Hence, our model is particularly relevant
when customer learning is an important driver of customer behavior
and the service environment is not observable to the customer. 

\subsection{Contributions to the Customers' Bayesian Learning Literature}

Our paper uses a Bayesian learning modeling framework. \citet{ching_invited_2013}
provide an extensive literature review and identifies four categories
of recent developments. Our work falls into the category: \textit{models
of correlated learning}. \citet{ching_invited_2013} summarize the
few existing studies in this stream (e.g., \citealt{erdem_empirical_1998,coscelli_empirical_2004,sridhar_investigating_2012,chan_treatment_2013})
and indicates that correlated learning means ``$\cdots$ learning
about a brand in one category by using the same brand in another category
and/or learning about one attribute (e.g., drug potency) from another
(e.g., side effects). This occurs if priors and/or signals are correlated
across products or attributes $\cdots$''. Previous literature has
explored the information spillover across multiple categories under
the same brand (e.g., \citealt{erdem_empirical_1998}) or type (e.g.,
\citealt{sridhar_investigating_2012}), or the information spillover
among attributes of a multi-attribute product (e.g., \citealt{coscelli_empirical_2004,chan_treatment_2013}).
Our study explores the information spillover from past shipping service
experiences, specifically the shipping experiences on multiple routes,
of the same logistics company. 

Our paper differs from previous studies of correlated learning in
several important aspects. First, we directly observe objective service
level metrics (e.g., the actual and planned time when the cargo is
delivered at the destination) while none of the previous studies observe
this information. Being able to observe service levels significantly
lowers our computational burden compared to previous studies, because
we can separate the estimation of a customer learning model from the
estimation of a purchase model, thus eliminating the need of integrating
out unobserved service quality. Therefore we are able to estimate
the learning process about the quality of $1000^{+}$ products, a
dramatically larger number compared to $2-7$ products usually seen
in the learning model literature, and obtain results within a reasonable
computational time. Moreover, having access to quality data enables
us to explore the potential impact of different quality measures (see
$\mathsection$4 for details), beyond the classic measures of perceived
quality mean and risk, on customers' purchase decisions. 

Second, the Bayesian hierarchical model provides a parsimonious and
structural (i.e., theory-based) approach to characterize the information
spillover process, and enables us to model the correlation among the
true qualities of the products. We note that previous studies of correlated
learning rely on estimating the covariance matrix of customers' quality
priors (e.g., \citealt{erdem_empirical_1998,coscelli_empirical_2004})
and/or the covariance matrix of noise in quality signals (e.g., \citealt{sridhar_investigating_2012,chan_treatment_2013,coscelli_empirical_2004}).
The number of parameters in these approaches rapidly (i.e., quadratically)
grows with the number of products for which spillovers are allowed.
In contrast, our model provides a more parsimonious approach to allow
for information spillovers among large numbers of products. Furthermore,
in our approach it is straightforward to allow for information spillovers
based on multiple product characteristics (e.g., distance and weight).
This can be easily modeled embedding a linear model of quality as
a function of product characteristics within the Bayesian learning
framework. Finally, and in contrast with previous approaches, because
we have access to quality data we can allow the true quality of the
same service (or product) to be different for different customers.
This is relevant in services, because a provider may set different
service levels for different customers (e.g., giving priority to high
volume customers). Note that without access to quality data it would
be very challenging to separately estimate heterogeneity in service
quality sensitivity and heterogeneity in true service quality. In
fact, to our knowledge all previous approaches have assumed that the
quality of a product or service is homogeneous across all customers. 

Third, we relax the constant experience variability and belief uncertainty
assumption that are used in the previous correlated learning literature.
Specifically, existing correlated learning models assume that consumers
are uncertain about the mean quality of a product or service, but
that the variability of the service (i.e., the precision of quality
signals) is constant and known to consumers (e.g., \citealt{coscelli_empirical_2004,sridhar_investigating_2012,chan_treatment_2013}).
The implication of this assumption is that as a consumer receives
more information, his uncertainty about the mean quality of the product
is lower (i.e., the consumer\textquoteright s perceived risk associated
with the product). In general, this is true when the new information
that a customer receives about a product is relatively congruent with
his or her prior knowledge of the product (e.g., when a customer has
experienced short delays in the past and her most recent experience
also involves a short delay). However, when this new information is
markedly inconsistent with the customer's prior beliefs, it could
raise rather than reduce the consumer's uncertainty associated with
the mean quality level. This is relevant for our application because
severe transport disruptions (e.g., a delay longer than a couple of
days) happen much more often than large quality changes for physical
goods. Accordingly, our model allows consumers to update their beliefs
about not only the mean product quality but also about the precision
of the quality signals. This feature yields a richer and more realistic
description of the customers' learning process. 

\subsection{Contributions to the Airline Customer Satisfaction Literature}

There is a large body of literature aiming to relate airline service
quality to microeconomic factors (e.g, price, competition, firm merging)
and another important stream that seeks to forecast flight delays
using statistical methods (refer to \citealt{deshpande_impact_2012}
for a detailed literature review). Different from our interests in
air cargo transport service, almost all of these studies have focused
on passenger airlines, with \citet{shang_exploiting_2016} as an exception,
in which the authors also studied air cargo transport delivery performance. 

There are a few studies sharing a similar interest to our paper in
terms of studying the customer-side consequences of airline transport
service performance focusing on passengers (e.g., \citealt{taylor_waiting_1994};
\citealt{forbes_effect_2008}; \citealt{anderson_impact_2008}). Our
paper contributes to this literature by providing evidence for the
air cargo industry, which suggests that customers and their beliefs
are more sensitive to ``poor'' rather than ``good'' service quality.
We also provide evidence that customers are averse to service experience
variability. 

\section{Empirical Setting \label{sec:data}}

We first describe the air cargo shipping process in $\mathsection$3.1.
In $\mathsection$3.2, we describe the main features of the data set
that was made available to us by a world leading freight forward company,
hereafter referred to as $AlphaShip$ (the true company name is disguised
for confidentiality). We collected additional information from Internet
sources, as described in $\mathsection$3.3. We conclude this section
in $\mathsection$3.4 with an exploratory analysis providing model-free
evidence of spillovers in customer learning. 

\subsection{Air Cargo Shipping Process}

An air cargo transport typically involves four parties: shippers (e.g.,
manufacturers), freight forwarders (forwarders in short), carriers
(i.e., airlines) and consignees (e.g., downstream manufacturers or
distributors). The shipping process starts with a request from the
shipper to the forwarder with certain shipping needs, such as origin
and destination cities, collection and delivery date, and cargo information
(pieces, weight and volume). The forwarder, who typically has reserved
spaces from airline partners, is often able to tell the shipper immediately
whether it has shipping space available that fits the customer's requirements;
and if so, the forwarder will provide the customer with a route map,
i.e., a shipping proposal that includes flight numbers and airline
information. The route map may be modified a few times before both
the shipper and the forwarder agree on a final version as well as
on the shipping fee. Then, the shipping process may start, which consists
of three stages: (1) the door to airport stage (D2A): the forwarder
picks up cargoes from the shipper at the required time, consolidates
cargoes sharing the same route if possible, and then sends cargoes
to the selected airline at an origin airport; (2) the airport to airport
stage (A2A): the airline is in charge of the cargoes until they arrive
at the destination airport; (3) the airport to door stage (A2D): the
forwarder accepts cargoes at the destination, and delivers them to
the consignees. In this process, it is the shipper who decides which
forwarder company to use and pays the shipping fees (in most of the
cases), so we refer to the shipper as the \textit{customer} in what
follows. 

As an intermediary, the forwarder is the service provider and direct
contact for its customers. Its responsibilities range from making
the first route map to updating the customer with alternative shipping
plans if the shipment does not proceed as planned (such as if the
flight took off before the cargoes could have been loaded).\footnote{Given that more than 90\% of the air cargo shipments are handled by
forwarders, we do not discuss the less frequent situations where the
airlines deal with air cargo shipments directly without an intermediate
forwarder. } As a result, if a shipment is not delivered on-time and hence customer
satisfaction drops, the forwarder faces the risk of losing the opportunity
of conducting business with this customer in the future. 

\subsection{Shipping Panel Data Descriptives}

The data provided by AlphaShip contains the records of its air freight
shipments from January to December in 2013. This data set follows
the Cargo 2000 standard, an air cargo industry information system
standard initiated in 1997 and currently widely used across the entire
industry (adopted by 80 major airlines, freight forwarders, ground
handler agents, trucking companies and IT providers). It records not
only the actual delivery time at each milestone, but also the planned
latest-by time for each milestone (see \citealt{shang_exploiting_2016}
for more details). By comparing the actual delivery time against the
planned delivery time, we can easily obtain the objective on-time
delivery performance for each shipment: whether the shipment is delayed
and how many hours it is delayed. For each shipment, the \emph{transport
delay}, is defined as the deviation of the actual final delivery time
at the destination from the planned delivery time:
\begin{eqnarray*}
\mbox{transport delay} & \equiv & \mbox{actual delivery time}-\mbox{planned delivery time}
\end{eqnarray*}
Neither earliness (i.e., a negative transport delay) nor tardiness
(i.e., a positive transport delay) may be desirable for customers.
While tardiness causes delays in production and/or product delivery
to all downstream customers, earliness may potentially yield additional
storage and handling costs. We use transport delay as the primary
measure of service quality of a shipment. 

The data set contains customer IDs, which allows us to track customer
purchases and model customers' learning about transport delay from
their experiences with AlphaShip. After data cleaning and selecting
customers with enough observations for model estimation (see Appendix
\ref{subsec:Data-Selection} for details), there are 725 customers
and 26,045 shipments left in the data. The cargoes are transported
from 53 countries to 153 countries on 2,897 routes all over the world.
Table \ref{Table: Summary Statistics} provides more information about
the data, including the choice predictors we use in our models (refer
to $\mathsection$4 for details) and presents customer level statistics.
\begin{table}[tb]
\centering{}\caption{Data Summary Statistics}
\label{Table: Summary Statistics}%
\begin{tabular}{>{\raggedright}p{3.2cm}>{\raggedleft}p{3cm}|>{\raggedright}p{4cm}>{\raggedleft}p{2.5cm}}
\hline 
\textsf{\textbf{\footnotesize{}Variable}} & \textsf{\textbf{\footnotesize{}Mean (std. dev)}} & \textsf{\textbf{\footnotesize{}Per customer Statistics}} & \textsf{\textbf{\footnotesize{}Mean (std. dev)}}\tabularnewline
\hline 
\textsf{\textit{\footnotesize{}Transport delay (hours)}} & \textsf{\footnotesize{}-0.77 (3.51)} & \textsf{\textit{\footnotesize{}Number of shipments}} & \textsf{\footnotesize{}41.83 (13.78)}\tabularnewline
\textsf{\textit{\footnotesize{}Chargeable weight (kg)}} & \textsf{\footnotesize{}1274.07 (4785.73)} & \textsf{\textit{\footnotesize{}Number of routes}} & \textsf{\footnotesize{}3.25 (1.00)}\tabularnewline
\textsf{\textit{\footnotesize{}Pieces}} & \textsf{\footnotesize{}1.13 (0.55)} &  & \tabularnewline
\textsf{\textit{\footnotesize{}Distance (km)}} & \textsf{\footnotesize{}7758.25 (3341.17)} &  & \tabularnewline
\hline 
\end{tabular}
\end{table}
As we can see from Table \ref{Table: Summary Statistics}, even though
the average transport delay is negative (i.e. cargoes delivered earlier
than planned time), the variance is significantly large, meaning that
the actual shipping service levels vary dramatically from shipment
to shipment. In particular, 26.0\% of the shipments are delayed and
9.2\% of the shipments are delayed more than 2 hours (regarded as
transport disruptions by industry standards, according to our discussion
with company executives).

\subsection{Price and Other External Data Descriptives}

Due to its sensitive nature, AlphaShip cannot provide us the shipping
fee associated with each shipment. In order to reconstruct prices,
we train a forecasting model by merging two external datasets. The
first is a sample of the official price data provided by IATA.\footnote{The IATA official price might be slightly different from the actual
shipping price, however both prices should be very similar.} The data contains price information on routes from six airports in
the Netherlands to almost all other airports in the world (846 airports),
spanning 12 months in 2013; in addition, the chargeable weight and
weight break associated with the prices are also included.\footnote{Chargeable weight = max\{volume weight (kg), actual weight (kg)\};
volume weight = volume (cubic meter)/6.} However, the number of routes included in the IATA sample is much
smaller than that in our data, so we seek additional data from a shipping
service quotation website \url{www.worldfreightrates.com}. Specifically,
we build a Python program to crawl shipping prices from the website
using the input (i.e. route, chargeable weight) extracted from our
Cargo 2000 data. Thus, we are able to obtain price information for
the majority of the routes and weight range. We crawled this website
twice, in September 2014 and February 2016 respectively, and each
process took approximately 15 days. Then, we merged the data from
the two sources to train the price forecasting model. The following
equation shows the best fitting forecasting model (adjusted $R^{2}=0.914$):
\begin{multline}
ln\left(P\right)=\gamma_{0}+\gamma_{1}\cdot Distance+\gamma_{2}\cdot Weight\\
+Weight\_break+Month+To\_country+Pieces\label{eq: Price Forecast}
\end{multline}
More details about the data and the estimation results can be found
in Table \ref{Table: Price Summaries}
\begin{table}[tb]
\centering{}\textsf{\footnotesize{}}%
\begin{minipage}[t]{0.43\columnwidth}%
\textsf{\footnotesize{}\caption{Price Data Descriptives}
\label{Table: Price Summaries}}{\footnotesize \par}
\begin{center}
\textsf{\footnotesize{}}%
\begin{tabular}{>{\raggedright}p{2.2cm}|>{\centering}p{0.8cm}>{\raggedleft}p{2.5cm}}
\hline 
 & \textsf{\footnotesize{}df} & \textsf{\footnotesize{}Mean (std. dev)}\tabularnewline
\hline 
\textsf{\footnotesize{}ln(Price) (\$)} & \textsf{\footnotesize{}1} & \textsf{\footnotesize{}6.7 (1.4)}\tabularnewline
\textsf{\footnotesize{}Distance (km)} & \textsf{\footnotesize{}1} & \textsf{\footnotesize{}7201.8 (3534.2)}\tabularnewline
\textsf{\footnotesize{}Weight (kg)} & \textsf{\footnotesize{}1} & \textsf{\footnotesize{}962.4 (5077.1)}\tabularnewline
\textsf{\footnotesize{}Weight\_break} & \textsf{\footnotesize{}23} & \tabularnewline
\textsf{\footnotesize{}Month} & \textsf{\footnotesize{}14} & \tabularnewline
\textsf{\footnotesize{}To\_country} & \textsf{\footnotesize{}220} & \tabularnewline
\textsf{\footnotesize{}Pieces} & \textsf{\footnotesize{}71} & \tabularnewline
\hline 
\end{tabular}
\par\end{center}{\footnotesize \par}%
\end{minipage}\textsf{\footnotesize{}\hfill{}}%
\begin{minipage}[t]{0.55\columnwidth}%
\begin{center}
\textsf{\footnotesize{}\caption{Price Forecast Model Estimates}
\label{Table: Price Estimates}}%
\begin{tabular}{>{\raggedright}p{2cm}|>{\centering}p{3.5cm}>{\raggedleft}p{1.5cm}}
\hline 
 & \textsf{\footnotesize{}Estimate (std. err)} & \textsf{\footnotesize{}t value}\tabularnewline
\hline 
\textsf{\footnotesize{}$\gamma_{0}$ (Intercept)} & \textsf{\footnotesize{}4.50 (7.69e-03){*}{*}{*}} & \textsf{\footnotesize{}584.90}\tabularnewline
\textsf{\footnotesize{}$\gamma_{1}$ (Distance)} & \textsf{\footnotesize{}1.15e-5 (3.26e-07){*}{*}{*}} & \textsf{\footnotesize{}35.32}\tabularnewline
\textsf{\footnotesize{}$\gamma_{2}$ (Weight)} & \textsf{\footnotesize{}1.65e-6 (1.65e-07){*}{*}{*}} & \textsf{\footnotesize{}10.00}\tabularnewline
\hline 
\multicolumn{2}{l}{\textsf{\footnotesize{}F-statistics (331, 473,520) = 1.62e+04}} & \tabularnewline
\multicolumn{3}{l}{\textsf{\footnotesize{}$R^{2}=0.914$; Adjusted $R^{2}=0.914$}}\tabularnewline
\hline 
\multicolumn{3}{l}{\textsf{\footnotesize{}$*p<0.05$; $**p<0.01$; $***p<0.001$.}}\tabularnewline
\end{tabular}
\par\end{center}%
\end{minipage}
\end{table}
 and Table \ref{Table: Price Estimates}. By using the coefficients
estimated from Equation (\ref{eq: Price Forecast}), we forecast the
price for each shipment obtained from AlphaShip. 

We note that the \textit{Distance} variable is not included in either
the Cargo 2000 data nor the IATA price sampler. Instead, we obtained
latitude and longitude information for all the airports in our data
set and calculated the great-circle distance between the original
and destination airport as the approximate distance for that route.
Although this great-circle distance is shorter than the actual flight
distance (especially if the shipping process includes multiple flights),
it proves to be useful not only in predicting shipping price but also
in specifying the learning and shipping choice model that we describe
next. 

\subsection{Exploratory Analysis}

In this subsection, we use simple descriptive models to explore how
customers use past experiences, especially experiences from similar
yet not identical services, to form quality beliefs which further
affect their future purchase decisions. For each customer $i$, we
denote by $r_{i}$ his most frequently used route during the 1-year
sample period. We then define as a dependent variable an indicator
$y_{it}$, where $y_{it}=1$ if customer $i$ ships on route $r_{i}$
during period $t$, and zero otherwise. We use the average transport
delay that customer $i$ has experienced on route $r_{i}$ until period
$t$ ($averag\_TR\_on\_this\_route$), and the average transport delay
customer $i$ has experienced on all other routes ($averag\_TR\_on\_other\_routes$)
as independent variables. We control for customer heterogeneity by
adding customer fixed effects (724 of them) and control for seasonality
by adding month fixed effects. Linear regression results are reported
in Table \ref{Table: Reduced-form}.
\begin{table}[tb]
\centering{}\caption{Linear Regression of Purchase Probability on Delays}
\label{Table: Reduced-form}\textsf{\footnotesize{}}%
\begin{tabular}{>{\raggedright}p{6cm}>{\centering}p{3.5cm}>{\raggedleft}p{1.2cm}}
\hline 
\textsf{\footnotesize{}Variable} & \textsf{\footnotesize{}Estimate (std. err)} & \textsf{\footnotesize{}z value}\tabularnewline
\hline 
\textsf{\footnotesize{}$averag\_TR\_on\_this\_route$} & \textsf{\footnotesize{}-9.92e-1 (3.89e-02){*}{*}} & \textsf{\footnotesize{}1.92}\tabularnewline
\textsf{\footnotesize{}$averag\_TR\_on\_other\_routes$} & \textsf{\footnotesize{}-9.65e-1 (1.72e-02){*}{*}{*}} & \textsf{\footnotesize{}1.98}\tabularnewline
\hline 
\multicolumn{3}{l}{\textsf{\footnotesize{}Pseudo $R^{2}=0.08$}}\tabularnewline
\multicolumn{3}{l}{\textsf{\footnotesize{}$*p<0.05$; $**p<0.01$; $***p<0.001$.}}\tabularnewline
\end{tabular}
\end{table}
As expected, higher average transport delays (e.g., more delays or
less earliness) on route $r_{i}$ lower customer $i$'s likelihood
of shipping on this route in the future. This provides evidence consistent
with customers forming quality beliefs using past experiences, which
in turn affect their future decisions. What is more relevant to our
research is that experiences on other routes also affect the customer's
purchase decisions on the focal route $r_{i}$. These results suggest
that when considering a focal route, experiences with other services
may spillover and affect future purchasing decisions on that focal
route. In the next section we will introduce a learning model that
takes into account these information spillovers. Moreover, this learning
model will allow us to consider a variety of policy simulation scenarios,
such as evaluating the impact of reducing not only the average magnitude
of delays, but also their variability. In addition, a learning model
will allow us to consider a service quality improvement on a focal
route and evaluate how quickly customers will change the likelihood
of shipping through not only the focal route, but also through other
routes. 

\section{Modeling Framework\label{sec:model}}

We formulate a model comprised of three interrelated components: demand
arrival, shipping choice decision and Bayesian learning about service
quality. We explain the details of each component in $\mathsection$4.1,
$\mathsection$4.2 and $\mathsection$4.3, respectively.

\subsection{Demand Arrival\label{subsec:Demand-Arrival}}

One limitation of our data is that it only includes the observed purchases
of shipping services performed by AlphaShip. In other words, we do
not observe data for cases where the customer contacted AlphaShip
to find out fees for shipping services but decided not to rely on
AlphaShip (i.e., using a competitor or simply not shipping during
that period). Similarly, our data does not separately identify periods
where a customer did not have a shipping need. Both scenarios from
the perspective of AlphaShip correspond to a no-purchase observation. 

In order to overcome this difficulty, we first borrow the modeling
approach used in \citet{newman_estimation_2014}. Specifically, we
divide the whole time horizon into a series of small discrete time
slices during which a customer arrival may or may not be observed
(an ``arrival'' means contacting AlphaShip for a shipping service
quote, which does not necessarily lead to a final purchase). The slices
must be sufficiently short so that the probability of two or more
shipping demands from one customer arriving during the same time period
is small. Arrivals are exogenous and modeled as a Bernoulli process
with parameter $\lambda_{i}$ corresponding to the probability that
customer $i$ arrives to AlphaShip in a given time period. In this
study, we choose half a week (i.e., 3.5 days) as a period. This choice
leads to a low probability of multiple arrivals.\footnote{For periods when multiple arrivals from the same customer are observed
(only 10.5\% among all the customer-period), we choose to keep the
route with the highest total arrival times from that customer.} 

A customer $i$ arriving at AlphaShip may be interested in shipping
on any route in his ``route set'', $\Upsilon_{i}\subset\triangle$,
where $\triangle$ is the set of all the routes (2000+ routes) in
the data set. For tractability, we focus on customers with a set containing
between 2 and 10 routes.\footnote{Customers with 1 or more than 10 routes represent 9\% and 12\%, respectively,
of the customer base. For more details about customer selection, please
refer to Appendix A1.} Furthermore, upon arrival the probability that he is interested in
shipping on route $j$ is given by $m_{ij}$, where $0\leq m_{ij}\leq1$
, $\sum_{j\in\Upsilon_{i}}m_{ij}=1$ and $m_{ij}=0$ for $j\in\triangle/\Upsilon_{i}$
(i.e., the choice of routes follows a multinomial distribution). 

Defining $d_{ijt}$ as an indicator which equals 1 if customer $i$
is interested in shipping on route $j$ in period $t$ and 0 otherwise,
then:
\begin{equation}
\mbox{Prob}\left(d_{ijt}=1\right)=\lambda_{i}m_{ij}.\label{eq: Demand Arrival}
\end{equation}

Note that this demand arrival process is independent from the actions
of AlphaShip. However, whether the customer decides to fulfill this
demand through AlphaShip will depend on AlphaShip's prices and prior
performance, as described next.

\subsection{Shipping Choice Model}

A customer $i$ with a shipping demand on route $j$ in period $t$
needs to decide whether to use AlphaShip or not (i.e., to use a competitor
or not to ship). The utility that customer $i$ derives from using
AlphaShip is given by (we use bold symbols for parameter vectors and
regular symbols for scalars):
\begin{eqnarray}
U_{ijt} & = & \beta_{i}^{0}+\beta^{p}\cdot Price_{jt}+\boldsymbol{\beta}^{X}\cdot X_{jt}+f\left(I_{it},\boldsymbol{\beta}_{i}^{q}\right)+e_{ijt},\label{eq: RUM}
\end{eqnarray}
while the utility of the outside option (i.e., not using AlphaShip)
is normalized to $U_{i0t}=e_{i0t}$. Here $e_{ijt}$ captures idiosyncratic
preferences of the customer unobserved to the researcher. In Equation
(\ref{eq: RUM}), we use an individual-level intercept $\beta_{i}^{0}$
to control for customer heterogeneity. We use the shipping price,
$\left(Price_{ijv}\right)$, as a factor influencing customers' utility,
which allows us to estimate customers' price sensitivity and to put
a dollar tag on the cost of service quality. Given that in most cases,
customers call the forwarder's local branch or use the forwarder's
online price quotation tool to get shipping prices before making purchase
decisions, we assume that prices are known to customers. However,
due to lack of pricing information, we use the price forecasted by
model (\ref{eq: Price Forecast}) in $\mathsection$3 to impute prices.
Furthermore, $X_{ijt}$ is a vector of controls which include: $(i)$
cargo chargeable weight ($Weight$); $(ii)$ monthly dummies (e.g.,
April, May) to control for seasonality across months and $\left(iii\right)$
a dummy variable equal to 1 when a period corresponds to the second
half of a week to control for seasonality within a week. One important
element in Equation (\ref{eq: RUM}) is the parametric function, $f\left(\cdot\right)$,
that captures the utility impact of a customer's beliefs about service
quality contained in his information set $I_{it}$. $\boldsymbol{\beta}_{i}^{q}$
is the function's parameter vector to be estimated. We will discuss
the functional form for $f\left(\cdot\right)$ in $\mathsection$4.3.3
after describing the customer service quality learning process in
$\mathsection$4.3.1 and $\mathsection$4.3.2. 

Assuming a standard extreme value distribution for $e_{ijt}$, the
random utility model described by Equation (\ref{eq: RUM}) becomes
a binary logit model. Let $y_{ijt}$ be an indicator that is set to
1 if consumer $i$ purchases shipping services on route $j$ at time
$t$, and 0 otherwise. Then the probability of a purchase, $y_{ijt}=1$,
given a demand $d_{ijt}$ is:
\begin{equation}
P_{ijt}\equiv\mbox{Prob}\left(y_{ijt}=1\mid d_{ijt}=1\right)=e^{V_{ijt}}/(1+e^{V_{ijt}}),\label{eq: Conditional Prob}
\end{equation}
where $V_{ijt}$ is the deterministic component of $U_{ijt}$: $V_{ijt}\equiv U_{ijt}-e_{ijt}$.
The model in Equation (\ref{eq: RUM}) includes not only a customer-specific
intercept, but also individual coefficients for the terms associated
with the effect of the shipping service quality $\left(\beta_{i}^{0},\boldsymbol{\beta}_{i}^{q}\right)$.
These individual coefficients are assumed to follow a multivariate
normal distribution with mean $\left(\beta^{0},\boldsymbol{\beta}^{q}\right)$
and for simplicity a diagonal covariance matrix $\Omega$, which we
seek to estimate from the data. 

\subsection{Customer Learning Spillover}

We now describe the customer learning process. This formulation will
allow for information spillovers under which a customers' experience
with one particular service may offer relevant information about the
service quality of other similar (yet not identical) services from
the same firm. Throughout, we model this learning process as being
independent across customers. Therefore, we describe the learning
mechanism for customer $i$, omitting the subscript $i$ in most of
the equations below for ease of exposition. We focus on transport
delay ($Q$) as the primary service quality measure. We refer to $Q$
as tardiness if $Q>0$ and earliness if $Q<0$. Given that the transport
process consists of a series of multiple stochastic events (i.e. multiple
connected flights, loading and unloading at the airports) and because
$Q$ can be either positive or negative, we assume that $Q$ follows
a normal distribution:
\begin{equation}
Q_{jt}\sim N(\mu_{jt},\sigma_{jt}^{2})\label{eq: True Distribution}
\end{equation}
where $\mu_{jt}$ and $\sigma_{jt}^{2}$ are the true mean and variance
of service quality on route $j$ for a customer's visit during period
$t$; $Q_{jt}$ is the customer's experienced service quality, which
he observes after the cargo has been delivered at the destination.
Although more flexible distributions could be used to model delays,
a normal distribution offers important tractability advantages. Also
note that different customers may have different true quality means
and variances (we have omitted the consumer subscript in $Q_{jt}$
and its moments for ease of exposition). We assume that customers
do not know the mean quality level $\mu_{jt}$ due to ``inherent
product variability'' \citep{ching_invited_2013} $\sigma_{jt}^{2}$,
which is referred to as ``experience variability'' in the learning
literature. Customers form beliefs about both the mean quality level
$\mu_{jt}$ and experience variability $\sigma_{jt}^{2}$ from the
signals $Q_{jt}$ received from their own usage experiences. In contrast
to previous empirical research in learning models, not only customers,
but also the researchers (us) observe the objective transport delay,
$Q_{jt}$. This data advantage enables us to explore more flexible
and general learning models and also allows us to relax strong assumptions
used in many previous literature, such as rational prior information
\citep{chan_treatment_2013} or full knowledge of experience variability
$\sigma_{jt}^{2}$ \citep{sridhar_investigating_2012}. We will explain
these in more details in $\mathsection$4.3.1 and $\mathsection$4.3.2.

\subsubsection{A Simple Bayesian Hierarchical Model for Customers' Learning}

First, we consider a simple Bayesian Hierarchical learning model (referred
to as the ``simple model'' in the following context) to characterize
customer learning spillovers. Throughout, we adopt the indexing convention
where subscript $t$ denotes the beginning of period $t$. As a simplification
of Equation (\ref{eq: True Distribution}), customers assume that
the true mean is constant across periods, i.e. $\mu_{jt}=\mu_{j}$
and hence it is only differentiated across routes $j$. Moreover,
we assume that the variance $\sigma_{jt}^{2}=\sigma^{2}$ is the same
for all service types and periods. Accordingly: 
\begin{equation}
Q_{jt}\sim N(\mu_{j},\sigma^{2})\label{eq: Simple Learning}
\end{equation}
Given that the service on all routes is arranged by the same logistics
firm, one might expect that the quality of the service for one route
might offer information about the quality of the service on another
route. Mathematically, a Bayesian hierarchical model allows for information
borrowing among the true qualities for different routes by adding
one more layer to the model. Here we assume that the true qualities
$\mu_{j},\forall j\in\Upsilon_{i}$, are generated exchangeably from
a common population, with a distribution governed by the hyper-parameters
$\mu$ and $\xi$: 
\begin{equation}
\mu_{j}\sim N\left(\mu,\xi^{2}\right),\forall j\in\Upsilon_{i}.\label{eq: Simple Hierarchy}
\end{equation}
The hyper-parameter $\mu$ provides the grand mean of qualities across
all routes, while $\xi^{2}$ measures the degree of heterogeneity
in service quality across routes. This heterogeneity parameter is
related to the degree of shrinkage of route qualities towards the
grand mean. Smaller values of this parameter imply that route qualities
are more similar and hence experiences for one route become more informative
about the quality of another route. 

We further allow these hyper-parameters ($\mu$ and $\xi^{2}$) to
be unknown to consumers. Hence, their prior beliefs about these parameters
are modeled using a Normal-Gamma distribution (this facilitates computations
due to its conjugacy properties). The consumer also doesn't know how
variable experiences are ($\sigma^{2}$), so an inverse gamma prior
is used to model consumer beliefs about this variability. Accordingly,
at the beginning of period $t=1$, customer $i$ has the following
initial beliefs: 
\begin{align}
\sigma^{2}\sim IG\left(\alpha_{\sigma},\delta_{\sigma}\right), &  & \mu\sim N\left(\mu_{0},\sigma_{\mu}^{2}\right), &  & \xi^{2}\sim IG\left(\alpha_{\xi},\delta_{\xi}\right)\label{eq: Simple Prior}
\end{align}
Here $IG(\alpha,\beta)$ represents the inverse-Gamma distribution
such that for $x\sim IG\left(\alpha,\beta\right)$ the pdf is $f\left(x\right)=\frac{\beta^{\alpha}}{\Gamma\left(\alpha\right)}x^{-\alpha-1}\mbox{exp}\left(-\beta/x\right)$
and the expectation is $E\left[x\right]=\frac{\beta}{\alpha-1}$.

Because our data presents a left truncation problem (i.e., most of
the customers have shipped with AlphaShip before our observation period
starts), we cannot assume the customers' priors to be the same during
the first period in our data. Borrowing a solution used in previous
studies (e.g., \citealt{mehta_role_2004} and \citealt{zhao_consumer_2011}),
we use the first 24 periods (around 3 months) of data as a pre-estimation
sample. We then assume that at the beginning of the pre-estimation
sample all customers have the same prior beliefs about experience
variability ($\sigma^{2}$), grand mean of product quality ($\mu$)
across routes and quality heterogeneity across routes ($\xi^{2}$),
such that $\sigma^{2}\sim IG\left(1.05,10\right)$, $\mu\sim N\left(0,30^{2}\right)$
and $\xi^{2}\sim IG\left(1.05,3\right)$. Here, contrary to the widely
used assumption of ``rational expectation of priors'' (e.g., \citealt{chan_treatment_2013}),
we use proper but relatively flat priors adding very weak assumptions
to the estimation. Based on this formulation, we calibrate individual
learning using the pre-estimation sample and calculate the posterior
distribution of $\{\sigma^{2},\mu,\xi^{2})\}$ for each customer at
the end of the 24\textsuperscript{th} period. Then we use these posteriors
as the corresponding prior for $\{\sigma^{2},\mu,\xi^{2})\}$ for
every customer in the 25\textsuperscript{th} period (see Table \ref{Table: Pre-estimation Priors}
in Appendix \ref{subsec: Priors} for details) and estimate the full
model using the remaining 80 periods in our data. 

The evolution of a customer's beliefs over time can then be updated
each period given every new piece of information acquired (i.e., shipping
experiences in our application) using Bayes' rule. Different from
the previous literature in which the posterior distribution of the
learning parameters has a closed-form expression, after adding a hierarchy
to our model to allow for learning spillovers (i.e., $\mu_{j}\sim N(\mu,\xi^{2})$),
a closed-form expression is not available. Thus we resort to Gibbs
sampling for calculating the posterior distribution of all quality
moments ($\mu_{j}$,$\sigma^{2}$,$\mu$ and $\xi^{2}$). A discussion
of the need of using a simulation-based method such as Gibbs sampling
and its advantages over other simulation methods such as Metropolis-Hastings
can be found in Appendix \ref{subsec:Advantages-of-Gibbs}. 

Accordingly, let $I_{t}$ denote the information set of a customer
at the beginning of time period $t$ ($25\le t\le T$). Then $I_{t}=$\{$Q_{j,p-1}$,
$\mu_{jp}^{E}$, $\sigma_{p}^{E}$, $\xi_{p}^{E}$, $\forall j\in\Upsilon_{i}$
and $25\le p\le t$\}, which contains both (i) the customer's shipping
experiences $Q_{j,p}$ in the past periods $1,..,t-1$; and (ii) his
expectation for shipping service quality at the beginning of each
period. Here, we use the $E$ superscript to denote a customer's ``estimated''
value for a quantity of interest (in our model, this estimated value
corresponds to the posterior mean). For example, $\mu_{jp}^{E}$ is
a customer's estimate of $\mu_{j}$ at the beginning of period $p$;
similar definitions apply to $\sigma_{p}^{E}$ and $\xi_{p}^{E}$.
The customer's shipping service quality estimates (e.g., $\mu_{jp}^{E}$)
are updated if he has new shipments delivered by AlphaShip in the
previous period; otherwise, the estimates are the same as before.
Assuming customer $i$ has shipments finished in period $t$, then
he updates his quality beliefs at the beginning of period $t+1$ using
the new information set $I_{t+1}$. A Gibbs sampler is implemented
to update customer beliefs by sampling each parameter from its full-conditional
posterior distribution until convergence (we use ``$\cdots$'' to
indicate all other parameters and data): 
\begin{eqnarray}
\mu_{j}\mid\cdots & \sim & N\left(\frac{n_{j}\xi^{2}\bar{Q}_{j}+\sigma^{2}\mu}{n_{j}\xi^{2}+\sigma^{2}},\;\frac{\xi^{2}\sigma^{2}}{n_{j}\xi^{2}+\sigma^{2}}\right),\label{eq: Gibbs Q}\\
\sigma^{2}\mid\cdots & \sim & IG\left(\alpha_{\sigma}+\frac{1}{2}\sum_{j\in\Upsilon_{i}}n_{j},\;\delta_{\sigma}+\frac{1}{2}\sum_{j\in\Upsilon_{i}}\sum_{p=1}^{t}y_{jp}^{*}\left(Q_{jp}-\mu_{j}\right)^{2}\right),\label{eq: Gibbs sigma}\\
\mu\mid\cdots & \sim & N\left(\frac{J\sigma_{\mu}^{2}\bar{\mu}+\xi^{2}\mu_{0}}{J\sigma_{\mu}^{2}+\xi^{2}},\;\frac{\sigma_{\mu}^{2}\xi^{2}}{J\sigma_{\mu}^{2}+\xi^{2}}\right),\label{eq: Gibbs mu}\\
\xi^{2}\mid\cdots & \sim & IG\left(\alpha_{\xi}+\frac{J}{2},\;\delta_{\xi}+\frac{1}{2}\sum_{j\in\Upsilon_{i}}\left(\mu_{j}-\mu\right)^{2}\right).\label{eq: Gibbs xi}
\end{eqnarray}
where the number of routes is denoted as $J\equiv\left|\Upsilon_{i}\right|$,
the average quality is given by $\bar{\mu}\equiv\frac{1}{J}\sum_{j\in\Upsilon_{i}}\mu_{j}$
; $y_{ijt}^{*}$ is an indicator that is set to 1 if customer $i$
had a shipment on route $j$ delivered during period $t$, and 0 otherwise;
$\bar{Q}_{j}=\frac{1}{n_{j}}\sum_{p=1}^{t}Q_{jp}y_{jp}^{*}$; and
$n_{j}=\sum_{p=1}^{t}y_{jp}^{*}$.\footnote{Note that $y_{ijt}^{*}$ is a shipment delivery indicator and $y_{ijt}$
($\mathsection$4.2) is a shipment start indicator. A shipment is
not always finished in the same period when it was initiated, thus
$y_{ijt}$ and $y_{ijt}^{*}$ are often different. } 

Given that Gibbs sampling is a simulation-based method, its outcome
is a set of values for all the parameters (i.e., $\mu_{j}$,$\sigma^{2}$,$\mu$
and $\xi^{2}$) drawn from their posterior distribution. To construct
estimates of mean quality beliefs such as $\mu_{j,t+1}^{E}$, we take
the average of the (posterior) draws for $\mu_{j}$, which gives an
estimate of the posterior mean of this parameter. The same approach
is used for all other parameters. The estimates, $\mu_{j,t+1}^{E}$,
$\mu_{t+1}^{E}$, $\sigma_{t+1}^{E}$ and $\xi_{t+1}^{E}$, are in
turn used as predictors in the quality function $f\left(I_{it},\boldsymbol{\beta}_{i}^{q}\right)$
that describes how quality beliefs affect a customer's decision to
rely on AlphaShip for shipping services (see $\mathsection$4.3.3).
\begin{figure}[tb]
\noindent \begin{centering}
\includegraphics[scale=0.5]{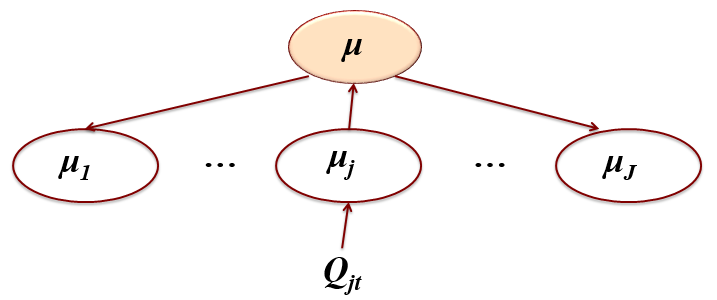}\caption{Spillover learning in the hierarchical Bayesian model.}
\label{Figure:Bayesiandiagram}
\par\end{centering}
\end{figure}

It is important to note that besides being an efficient computational
tool to estimate the posterior distribution of quality beliefs, the
Gibbs sampling steps are also useful to understand how learning spillovers
shape these beliefs. Specifically, a new quality (i.e., delay) observation
on route $j$ ($Q_{jt}$) changes the customer's beliefs about the
mean transport delay on route $j$ ($\mu_{j}$, see Figure \ref{Figure:Bayesiandiagram}
and equation \ref{eq: Gibbs Q}). This in turn influences the customer's
(grand) mean quality ($\mu$) about all routes served by AlphaShip
(see equation \ref{eq: Gibbs mu}). Through this iterative process,
the new beliefs about service quality grand mean shifts the beliefs
of all other routes ($\mu_{-j}$) and more strongly when the degree
of route quality heterogeneity $\xi^{2}$ is small (see equation \ref{eq: Gibbs Q}).
Consequently, information about one product (route) is used not only
to update beliefs about that product, but also about others. 

Furthermore, new experiences not only change consumer beliefs about
the mean qualities ($\mu_{j}$ and $\mu$), but also about the variability
of the experiences. More specifically, experience variability $\left(\sigma^{2}\right)_{t}^{E}$
reflects the variance of service quality $Q_{jt}$ around its mean
($\mu_{j}$). The properties of the inverse-gamma distribution imply
that the consumer's estimate of experience variability $\left(\sigma^{2}\right)_{p}^{E}$
can be expressed as: 
\begin{equation}
E\left[\sigma^{2}\mid\cdots\right]=\left[\delta_{\sigma}+\frac{1}{2}\sum_{j\in\Upsilon_{i}}\sum_{p=1}^{t}y_{jp}^{*}\left(Q_{jp}-\mu_{j}\right)^{2}\right]/\left(\alpha_{\sigma}-1+\frac{1}{2}\sum_{j\in\Upsilon_{i}}n_{j}\right)\label{eq: Variance Expectation}
\end{equation}
Note that the denominator in this equation increases with the arrival
of new information, because $\sum_{j\in\Upsilon_{i}}n_{j}$ is the
cumulative number of experiences. However, the corresponding numerator
also increases when new information arrives. In particular, if the
new delay information $Q_{jp}$ is very inconsistent with mean quality
beliefs $\mu_{j}$ (e.g., after experiencing a unusually long delay),
beliefs about experience variability will increase dramatically. 

It is important to note that the standard learning models in the literature
assume experience variability $\sigma^{2}$ to be a constant known
by the customer. In addition, the variance of prior (i.e., initial)
beliefs about $\mu_{j}$ is also assumed to be constant (e.g., consumers
might start with very vague beliefs). This last assumption is reasonable
when modeling the learning process about the quality (or other attributes)
of a single product in isolation. However, this assumption also makes
beliefs about the mean quality of a product to monotonically become
more precise as the customer gains new experiences. \citet{zhao_consumer_2011}
relax this assumption by assuming an inverse-gamma distribution for
belief uncertainty. Our model not only relaxes the assumption about
the experience variability $\sigma^{2}$ being known to the consumer,
but also allows the uncertainty about mean quality beliefs to potentially
increase or decrease as new information is gained by the customer.
Accordingly, to the best of our knowledge our paper is the first study
in correlated learning to relax both assumptions, thus providing a
more flexible quality learning structure. 

A similar pattern is observed when focusing on the evolution of beliefs
about route quality heterogeneity $\left(\xi^{2}\right)_{t}^{E}$,
which measures service quality differences across routes. When new
information about a route is dramatically different from prior beliefs
(e.g., an unusually early delivery), this will shift mean quality
beliefs about that route, and if doing so the estimated quality of
that route ($\mu_{jt}^{E}$) shifts away from AlphaShip's grand mean
quality ($\mu$), then the estimated route heterogeneity $\left(\xi^{2}\right)_{t}^{E}$
will increase. 

To help readers better understand the information spillover process,
we estimate the learning process for a customer (in our data) who
operates on three routes denoted by $\{a$, $b$, $c\}$. We plot
the changes to his estimated mean quality beliefs for each route $\mu_{at}^{E}$,
$\mu_{bt}^{E}$, $\mu_{ct}^{E}$, grand mean quality $\mu_{t}^{E}$,
experience variability $\sigma_{t}^{E}$ and quality heterogeneity
across routes $\xi_{t}^{E}$ throughout the observation period. As
we can see from the top panel in Figure
\begin{figure}[tb]
\begin{centering}
\noindent\begin{minipage}[t]{1\columnwidth}%
\begin{center}
\includegraphics[width=0.8\columnwidth]{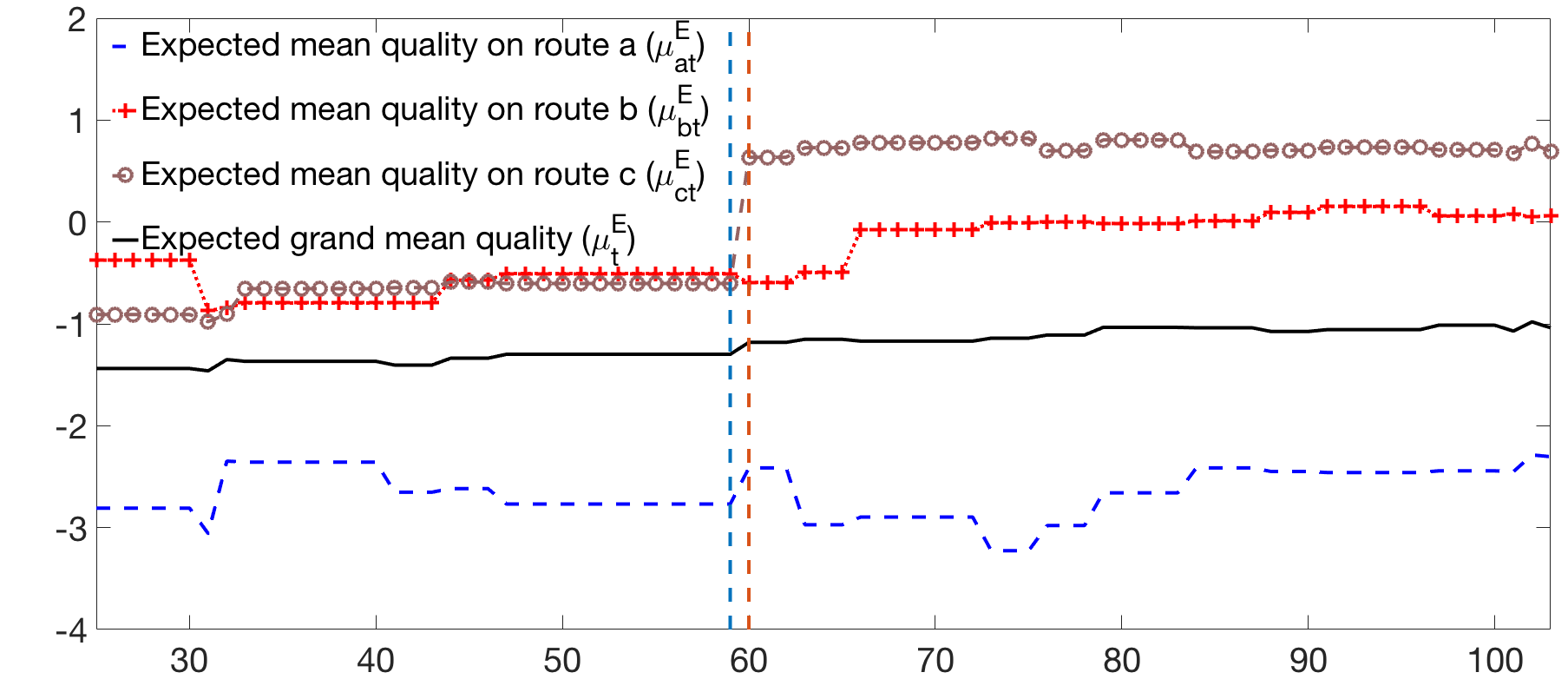}
\par\end{center}%
\end{minipage}
\par\end{centering}
\begin{centering}
\noindent\begin{minipage}[t]{1\columnwidth}%
\begin{center}
\includegraphics[width=0.8\columnwidth]{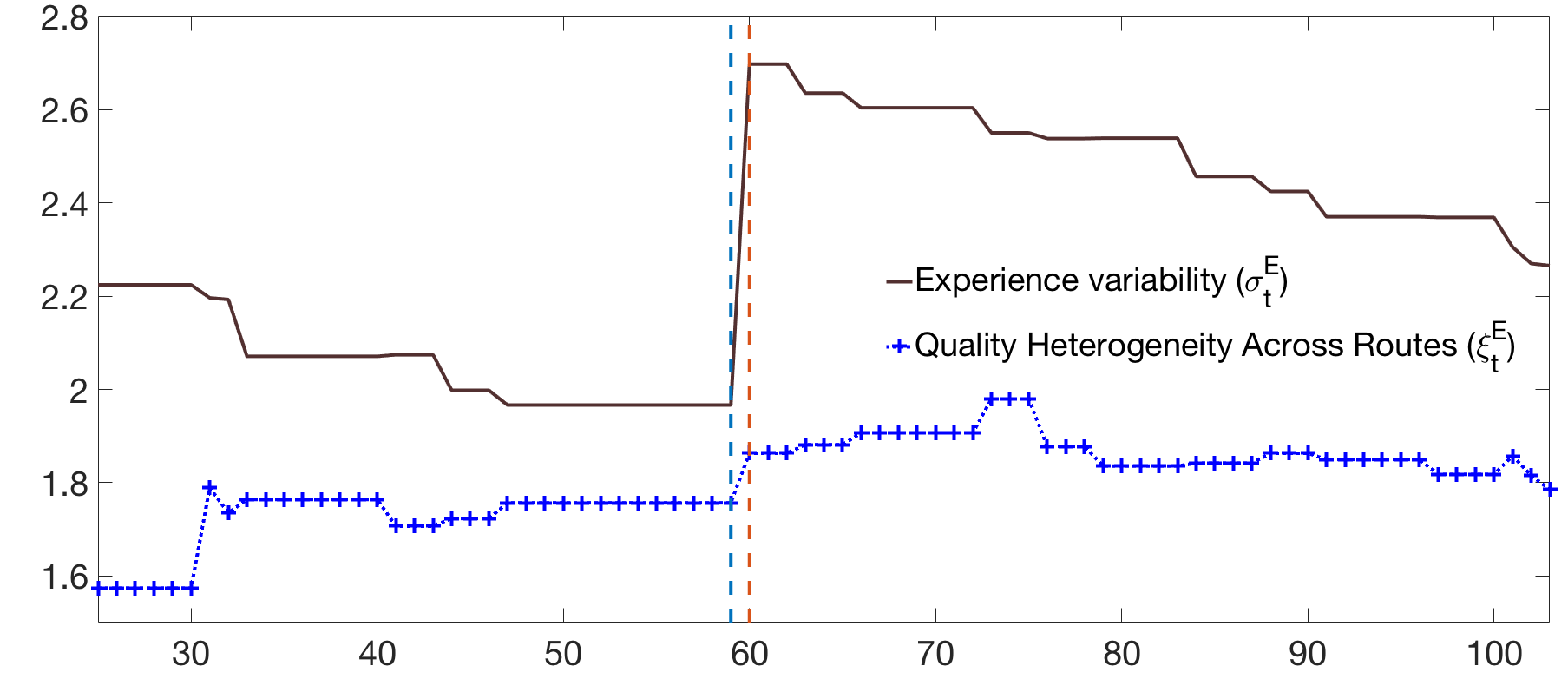}
\par\end{center}%
\end{minipage}
\par\end{centering}
\caption{Top: updates of perceived shipping quality on each route and overall
shipping quality; Bottom: updates of experience variability and belief
uncertainty}
\label{Figure: Period updates}
\end{figure}
 \ref{Figure: Period updates}, the estimated mean quality for each
route changes over time and so does the grand mean quality ($\mu_{t}^{E}$)
which takes values between the highest and lowest of the route quality
beliefs. To maintain the readability of the figure, we did not plot
the quality of all shipping experiences $Q_{ijt}$. Instead, let us
focus on the events corresponding to $t=59$. We use two vertical
lines in the top and bottom panels to highlight the period in which
this ``experience'' was gained ($t=59$) and the following period
($t=60$). The top panel shows a large increase of estimated quality
(i.e., delay) on route $c$ in period 60. This is caused by a severe
delay of almost 9 hours on this route in period 59. This severe delay
also causes the estimated grand mean quality $\mu_{t}^{E}$ to increase
in period 60 (see the top plot). Furthermore, as we explained before,
the large inconsistency between the experience in period 59 and the
estimated mean quality $\mu_{c59}^{E}$ causes a surge in experience
variability $\sigma_{t}^{E}$ following the severe transport disruption
as depicted in the bottom plot. 

With this large increase in experience variability $\sigma_{t}^{E}$
and a relatively small increase in quality heterogeneity across routes
$\xi_{t}^{E}$, the estimated mean quality for each route $j$ (i.e.,
$\mu_{jt}^{E}$) places more weight on the grand mean quality $\mu_{t}^{E}$
compared to the average observed quality on that route ($\bar{Q}_{j}$).
Mathematically, this can be seen using the Gibbs sampling formula
$E\left[\mu_{j}\mid\cdots\right]=\frac{n_{j}\xi^{2}\bar{Q}_{j}+\sigma^{2}\mu}{n_{j}\xi^{2}+\sigma^{2}}$
(Equation (\ref{eq: Gibbs Q})). Note that this expectation is a weighted
average between the average observed quality on route $j$ ($\bar{Q}_{j}$)
and the overall mean quality across all routes ($\mu$). As a result,
when quality exhibits large variability ($\sigma_{t}^{E}$), the estimated
mean qualities on routes $a$ and $b$, $\mu_{at}^{E}$ $\mu_{bt}^{E}$,
respectively, more strongly shift towards the overall quality $\mu_{t}^{E}$. 

\subsubsection{Bayesian Regression Hierarchical Model for Customers' Learning}

We now consider an alternative setting where customers learn from
their shipping experiences at AlphaShip taking into account the characteristics
of different routes when using information about one route to learn
about the quality on another route. For example, when forming beliefs
about the quality of a route, the quality of more similar routes (e.g.,
routes sharing an origin or destination airport) might be more informative.
Formally, this approach is based on embedding a Bayesian hierarchical
regression within the learning model. Accordingly, instead of modeling
consumer beliefs in a given period $t$ for route $j$ as $\mu_{jt}=\mu_{j}$
in Equation (\ref{eq: True Distribution}), we allow $\mu_{jt}$ to
not only depend on the identity of that route as in $\mathsection$4.3.1
but also with the specific characteristics of that route. Focusing
on distance as one of those characteristics, $\mu_{jt}$ becomes a
linear function of route distance:
\begin{eqnarray}
\mu_{jt} & = & \theta_{j}+\gamma\cdot Distance_{j},\label{eq: Regress Hierarchy}
\end{eqnarray}
where $\theta_{j}$ is a route-level mean quality intercept for route
$j$; $Distance_{j}$ is the great-circle distance of route $j$;
and $\gamma$ measures the effect of distance on the mean transport
delay. Due to data constraints, especially the missing no-purchase
data as explained in $\mathsection$4.1, we use only distance as a
predictor of mean quality. However, our model and estimation method
can be easily extended to include more route characteristics if more
observations were available. As before, we add a hierarchy to these
parameters to allow for information sharing: 
\begin{align*}
\theta_{j}\sim N\left(\mu,\xi^{2}\right), & \forall j\in\Upsilon_{i},
\end{align*}
and we adopt a Normal-Gamma hyper-prior distribution: $\theta\sim N\left(\mu_{0},\sigma_{\mu}^{2}\right)$,
$\gamma\sim N\left(\gamma_{0},\sigma_{\gamma}^{2}\right)$, $\xi^{2}\sim IG\left(\alpha_{\xi},\delta_{\xi}\right)$,
$\sigma^{2}\sim IG\left(\alpha_{\sigma},\delta_{\sigma}\right)$.
We solve the left truncation problem in the same way as before by
using the first 24 periods data as pre-estimation sample and adopting
the same initial beliefs except for the distance coefficient for which
we assume a vague prior: $\gamma\sim N\left(0,30^{2}\right)$. Finally,
under this learning model each customer's information set becomes
$I_{t}=$$\{$$Q_{j,p-1}$, $\mu_{jp}^{E}$, $\theta_{jp}^{E}$, $\gamma_{p}^{E}$,
$\sigma_{p}^{E}$, $\xi_{p}^{E}$ $\forall$$j\in\Upsilon_{i}$ and
$25\le p\le t$$\}$, $\forall25\le t\le T$. For detailed Gibbs sampling
formulas, please refer to Appendix \ref{subsec:Gibbs-Sampling-of}
and \ref{subsec:Computation}. 

\subsubsection{Service Quality Related Factors in the Utility Function}

In this subsection we discuss how shipping service quality beliefs
may affect customers' shipping decisions through $f\left(I_{jt},\boldsymbol{\beta}^{q}\right)$
in Equation (\ref{eq: RUM}) (subscript $i$ is omitted for ease of
exposition). In particular we focus on three characteristics of this
function. 

First, we consider the potentially asymmetric effect of the estimated
mean quality $\mu_{jt}^{E}$. In our application positive and negative
quality values have a different meaning, they imply tardiness and
earliness, respectively and this different meaning may yield different
consequences for customers. Therefore, we will allow positive and
negative values of the mean quality to have a different impact on
the utility function that determines a customer's shipping decisions. 

Second, we explore the possible nonlinear effect of the estimated
mean quality (i.e., transport delay). Customers might respond differently
to minor delays (e.g., 2 hours) than to disruptive delays (e.g., 2
days). For example, compared to the smaller inconvenience brought
by a 2-hour day, a delay of 2 days may probably disrupt the customer's
production or delivery plan and cause more severe loss of business.
Accordingly, we consider a quadratic function of the estimated mean
quality $\mu_{jt}^{E}$ to account for this possibility. 

Finally, similar to several previous learning models in the literature
(e.g., \citealt{erdem_empirical_1998,sridhar_investigating_2012}),
we consider that customers might be risk averse. The uncertainty about
service quality $Q_{j}$ has two components: $Var\left[Q_{j}\right]=$$E\left[\left(\sigma^{2}\right)_{jt}\right]+Var\left[\mu_{jt}\right]$.
The first component $E\left[\left(\sigma^{2}\right)_{jt}\right]$
measures the estimated degree of experience variability and originates
from inherent variability in the delivery of the service and thus
cannot be reduced with additional experiences. The second component
$Var\left[\mu_{jt}\right]$ measures customers' uncertainty about
the true value of the mean quality and it originates from a customer's
limited knowledge. The effect of these two uncertainty components
on customer utility may differ. Previous studies have explored customers'
risk aversion by either combining the two components (i.e., estimating
the effect of the total uncertainty $Var\left[Q_{j}\right]$, such
as \citet{erdem_empirical_1998}) or only focusing on the effect of
the uncertainty associated with learning the true value of the mean
quality $Var\left[\mu_{jt}^{E}\right]$ (e.g., \citealt{coscelli_empirical_2004,sridhar_investigating_2012,zhao_consumer_2011}).
In contrast, in our model we separately estimate and compare the effect
of these two sources of uncertainty. 

\subsection{Model Identification}

Given that customers directly observe the service quality $Q_{ijt}$
for each of their shipments with AlphaShip, the identification of
the learning model (see $\mathsection$4.3) is a standard Bayesian
statistics problem: the parameters will always have proper posterior
distributions given the prior distributions that we use, and the magnitude
of the posterior variance depends on the information contained in
the Cargo 2000 shipping data. So we focus on the identification of
the demand arrival (see $\mathsection$4.1) and shipping choice models
(see $\mathsection$4.2). 

From Equation (\ref{eq: Demand Arrival}) and (\ref{eq: Conditional Prob}),
we obtain the probability of observing a purchase from customer $i$
on route $j$ at time $t$: 
\[
P\left(y_{ijt}=1\right)=P\left(y_{ijt}=1\mid d_{ijt}=1\right)\cdot P\left(d_{ijt}=1\right)=\lambda_{i}m_{ij}\cdot e^{V_{ijt}}/(1+e^{V_{ijt}})
\]
In the above equation, the parameter $\lambda_{i}$ determines the
overall demand level of customer $i$, given that $P\left(y_{ijt}=1\right)\propto\lambda_{i}$.
However, the customer-level intercept $\beta_{i}^{0}$ also plays
the role of adjusting the overall probability of shipping: if $\beta_{i}^{0}$
increases then $e^{V_{ijt}}/(1+e^{V_{ijt}})$ increases. As a result,
separately identifying both $\lambda_{i}$ and $\beta_{i}^{0}$ is
very difficult, especially when having few observations (around 20
for each customer). For example, the data generated from a high $\lambda_{i}$
and low $\beta_{i}^{0}$ customer can be very similar to that from
a customer with low $\lambda_{i}$ and high $\beta_{i}^{0}$. As a
result, we set $\lambda_{i}=1$ for all customers and focus on estimating
$\beta_{i}^{0}$. 

Now consider $m_{ij}$, which is a parameter of a multinomial distribution.
These parameters must satisfy the following constraints $0<m_{ij}\le1$
for $j\in\Upsilon_{i}$ and $\sum_{j\in\Upsilon_{i}}m_{ij}=1$ for
customer $i$. To enforce these constraints, we estimate $\bar{m}_{ij}$
instead of $m_{ij}$ , where $m_{ij}\equiv exp\left(\bar{m}_{ij}\right)/\sum_{j\in\Upsilon_{i}}exp\left(\bar{m}_{ij}\right)$
and $-\infty<\bar{m}_{ij}<\infty$. Without loss of generality, we
normalize the value of $\bar{m}_{ij}$ for the first route ever chosen
by customer $i$ to 0. The identification of the remaining model parameters
(i.e. $\beta^{p}$) is standard in discrete choice models and will
be omitted for brevity. 

Finally, we conducted a Monte Carlo simulation study to test our methodology.
We simulated the shipping behavior of 80 customers for 100 periods.
The arguments, such as shipping service quality, used in the simulation
test are set close to those in the real data. Results show that our
parameters can be reasonably recovered from the simulated data. These
results provide evidence that our estimation approach can recover
the true parameters of the data generating process (detailed results
are available in the online appendix \ref{subsec:Numerical-Experiment}).
For more details about the estimation, please refer to Appendix \ref{subsec:Missing-Data-Interpolation}
and \ref{subsec:Shipping-Choice-Model}. 

\section{Results and Discussion\label{sec:Results-and-Discussion}}

We first describe different learning models and compare their goodness-of-fit
in $\mathsection$5.1. In $\mathsection$5.2, we then provide the
results of the demand arrival and shipping choice model, particularly,
we provide results for the choice models based on different learning
assumptions. 

\subsection{Quality Learning Model}

After every experience customers in our study obtain an objective
measure of service quality (i.e., transport delay). Different from
most models in the empirical quality learning literature, because
we also observe these service quality metrics, the estimation of each
customers' learning process can be performed independently from the
estimation of the remaining model parameters. Before presenting the
estimates of the learning model, we first compare the goodness-of-fit
of several competing learning models using the shipping service quality
data. 

\subsubsection{Learning Model Comparison}

In addition to the Bayesian hierarchical model explained in section
4.1, we further consider several benchmark learning models. The first
benchmark is a short-memory learning model under which customers only
rely on the most recent experience when anticipating the service quality
they will receive. Specifically, we let 
\begin{equation}
f\left(I_{t},\boldsymbol{\beta}^{q}\right)=Q_{ijt-1}\label{eq: Benchmark 1}
\end{equation}
where $Q_{ij,t-1}$ is the actual shipping service quality experienced
by customer $i$ on route $j$ in period $t-1$ ($Q_{ij,t-1}$ is
the same as $Q_{ij,t-2}$ if there is no new experience in period
$t-1$), and is set to 0 if the customer has not chosen route $j$
by period $t$ to rely on the services of AlphaShip. 

The second benchmark model describes an independent learning process
\textemdash{} a customer only updates his beliefs about shipping quality
on a route using the usage experience on that specific route, without
borrowing information from other routes he has experienced before.
Recall that we model $Q$ using a normal distribution, i.e. $Q_{jt}\sim N\left(\bar{Q}_{jt},\sigma_{jt}^{2}\right)$
(see Equation (\ref{eq: True Distribution}) in $\mathsection$4.3.1).
In the independent learning model, customer beliefs about the mean
quality $\mu_{jt}$ can be expressed as follows: 
\begin{align}
\mu_{jt}\sim N\left(\mu_{0j},\xi_{j}^{2}\right)\label{eq: Benchmark 2}
\end{align}
Due to the lack of data for each customer on each route, we do not
consider an independent learning model with predictors here. 

The third benchmark model relies on full information pooling \textemdash{}
for each customer the information coming from all routes is equally
informative to learn about the mean quality of a particular route.
Specifically, we consider two sub-models: benchmark 3A\footnote{In both benchmark model 2 and 3, we use the first 24 period pre-estimation
sample to obtain the customer-level priors for the rest of the periods} (equation (\ref{eq: Benchmark 3A})) and benchmark 3B (equation (\ref{eq: Benchmark 3B})),
where model 3B has an additional predictor $Distance$ compared to
model 3A:
\begin{align}
 & \mu_{jt}=\mu, &  & \mu\sim N\left(\mu_{0},\xi^{2}\right)\label{eq: Benchmark 3A}\\
 & \mu_{jt}=\theta+\gamma\cdot Distance, &  & \theta\sim N\left(\mu_{0},\xi^{2}\right), &  & \gamma\sim N\left(\gamma_{0},\sigma_{\gamma}^{2}\right)\label{eq: Benchmark 3B}
\end{align}
We fit each alternative learning model using the quality (i.e., delay)
data. Table 
\begin{table}[tb]
\begin{centering}
\textsf{\footnotesize{}\caption{Model Comparison and Goodness of Fit}
\label{Table: Benchmark Models}}
\par\end{centering}{\footnotesize \par}
\centering{}\textsf{\small{}}%
\begin{tabular}{>{\raggedright}p{1.5cm}>{\centering}p{2.4cm}>{\centering}p{2.4cm}>{\centering}p{2.4cm}>{\centering}p{2.4cm}>{\raggedleft}p{2.4cm}}
\hline 
 & \textsf{\footnotesize{}Benchmark Model 2} & \textsf{\footnotesize{}Benchmark Model 3A} & \textsf{\footnotesize{}Benchmark Model 3B} & \textsf{\footnotesize{}Simple hierarchical model} & \textsf{\footnotesize{}Full model}\tabularnewline
\hline 
\textsf{\footnotesize{}Features} & \textsf{\footnotesize{}Independent learning} & \textsf{\footnotesize{}Information pooling (IP)} & \textsf{\footnotesize{}Regression IP} & \textsf{\footnotesize{}Simple hierarchical} & \textsf{\footnotesize{}Regression hierarchical}\tabularnewline
\textsf{\footnotesize{}Model} & \textsf{\footnotesize{}Eq (\ref{eq: Benchmark 2})} & \textsf{\footnotesize{}Eq (\ref{eq: Benchmark 3A})} & \textsf{\footnotesize{}Eq (\ref{eq: Benchmark 3B})} & \textsf{\footnotesize{}$\mathsection4.3.1$} & \textsf{\footnotesize{}$\mathsection4.3.2$}\tabularnewline
\textsf{\footnotesize{}$-LL$} & \textsf{\footnotesize{}$4.540e+4$} & \textsf{\footnotesize{}$4.559e+4$} & \textsf{\footnotesize{}$4.659e+4$} & \textsf{\footnotesize{}$4.384e+4$} & \textsf{\footnotesize{}$4.542e+4$}\tabularnewline
\textsf{\footnotesize{}$DIC$} & \textsf{\footnotesize{}$1.219e+5$} & \textsf{\footnotesize{}$1.208e+5$} & \textsf{\footnotesize{}$1.236e+5$} & \textsf{\footnotesize{}$1.179e+5$} & \textsf{\footnotesize{}$1.223e+5$}\tabularnewline
\hline 
\end{tabular}{\small \par}
\end{table}
\ref{Table: Benchmark Models} provides fit measures: negative log-likelihood
($-LL$) and Deviance Information Criterion ($DIC$) of the benchmark
models, the simple hierarchical model (see $\mathsection$4.3.1) and
the regression hierarchical model (see $\mathsection$4.3.2).\footnote{DIC is a hierarchical modeling generalization of the AIC (Akaike information
criterion) and BIC (Bayesian information criterion, and is particularly
useful in Bayesian model selection.} For both $-LL$ and $DIC$, the smaller the values, the better the
model fits the shipping experience data. Given that benchmark model
1 does not involve learning, it is not included in Table \ref{Table: Benchmark Models}.
Among all the models, the simple hierarchical model provides the best
fit in terms of both $-LL$ and $DIC$. This implies that a customer
that relies on this learning model more accurately estimates the service
quality of a future experience. However, it is important to note that
the model that provides the best fit to the shipping quality data
may not necessarily be the actual learning model that customers use.
We will explore this issue in $\mathsection$5.2 where we will compare
the ability of the different learning models to provide useful predictors
of customer choices.

In terms of the fit of the quality data, a closer look into Table
\ref{Table: Benchmark Models} reveals that adding the $Distance$
predictor decreases model fit, and this is shown by the fact that
the corresponding model \textit{without} this predictor performs better
for both the hierarchical model (simple hierarchical model vs. full
model) and the information pooling models (models 3A vs. 3B). Contrary
to a linear regression where adding more predictors does not decrease
in-sample fit, in a learning model this is not necessarily the case.
In a learning model the importance of a predictor ($\gamma$ in our
model) changes over time based on the gained experiences. So all previous
experiences determine the weight of a predictor, however this weight
is then used to predict a new experience.\footnote{To illustrate ideas, suppose that all previous experiences with long
distance shipments were satisfactory (no delays), while those with
short distances were not. Then this will give a strong weight to distance
when the consumer anticipates the quality of future experiences. However,
if the next experiences reverse this pattern (e.g., they involve a
satisfactory short distance shipment and an unsatisfactory long distance
shipment), then using distance will not make quality beliefs more
accurate for these new experiences. } 

Furthermore, the hierarchical model systematically fits the data better
than the information pooling model. Finally we observe that the fit
of the independent learning model is similar to that of the information
pooling model 3A. Overall, the model comparison results justify our
interest of considering spillovers in our learning model specifications.
Before examining the ability of each learning model to predict customer
choices, we discuss in more detail the features of the best-fitting
learning model in the next subsection. 

\subsubsection{Estimation of the Bayesian Simple Hierarchical Model}

Given that the simple hierarchical model (see $\mathsection$4.3.1)
provides the best fit of the shipping quality data, in what follows
we discuss the results for this model, while the results of other
learning models are similar. 

Under this model, each customer learns from his own experiences without
using information from other customers. Therefore there are multiple
parameters $\{\mu_{ij}$, $\mu_{i}$, $\sigma_{i}^{2}$, $\xi_{i}^{2}$,
$\forall j\in\Upsilon_{i}$, $i=$$1$, $\cdots$, $N\}$ to estimate
for each customer, leading to more than 2,000 parameters in total.
Due to the large number of individual-level parameters, in Table 
\begin{table}[tb]
\begin{centering}
\caption{Learning model parameters' population mean and 90\% quantile interval}
\label{Table: Bayesian Estimates}
\par\end{centering}
\centering{}%
\begin{tabular}{>{\raggedright}m{1cm}>{\centering}m{2cm}>{\raggedleft}m{3.3cm}}
\hline 
 & \textsf{\footnotesize{}Mean} & \textsf{\footnotesize{}Percentiles (5\%, 95\%)}\tabularnewline
\hline 
\textsf{\textbf{\footnotesize{}$\mu_{iT}^{E}$}} & \textsf{\footnotesize{}-0.727} & \textsf{\footnotesize{}(-2.863, 1.560)}\tabularnewline
\textsf{\textbf{\footnotesize{}$\sigma_{iT}^{E}$}} & \textsf{\footnotesize{}2.592} & \textsf{\footnotesize{}(1.150, 5.122)}\tabularnewline
\textsf{\footnotesize{}$\xi_{iT}^{E}$} & \textsf{\footnotesize{}1.759} & \textsf{\footnotesize{}(1.280, 2.572)}\tabularnewline
\hline 
\end{tabular}
\end{table}
 \ref{Table: Bayesian Estimates}, we only provide the population
mean and 90\% interval of the grand mean quality $\mu_{iT}^{E}$,
experience variability $\left(\sigma^{2}\right)_{iT}^{E}$ and route
quality heterogeneity $\left(\xi^{2}\right)_{iT}^{E}$ as estimated
by customers under this model during the last period $T$ in our dataset.
The summary statistics of the route-specific quality beliefs, $\mu_{ijT}^{E}$,
are not included in Table \ref{Table: Bayesian Estimates} due to
the varying set of routes $\Upsilon_{i}$, across customers. 

From the results in \ref{Table: Bayesian Estimates}, we observe considerable
heterogeneity among customers with a 90\% interval implying that some
customers anticipate mean delays of 1.56 hours on one extreme, while
others expect mean earliness of 2.86 hours on the other extreme. In
previous learning models, the heterogeneity in customers' learning
processes mainly comes from the limited number of trials (e.g., \citealt{coscelli_empirical_2004}).
Due to experience variability, customers can only observe noisy signals
of the true quality. When the number of experiences is small, customers
may form beliefs which are far from the true mean quality, thus resulting
in different service quality beliefs. In contrast to previous approaches
in the learning literature, we do not restrict the underlying true
mean quality $\mu_{ijt}$ to be the same for all customers. 

In addition, we find that the population mean of $\mu_{iT}^{E}$ equals
-0.727 is close to the empirical shipping quality average of -0.77
hours (see Table \ref{Table: Summary Statistics}), and the population
mean of the quality standard deviation $\sigma_{iT}^{E}$ equals 2.56
and is somewhat smaller than the empirical standard deviation of shipping
quality 3.51. This shows that after one year and under this learning
model, customer quality beliefs are on average close to the true empirical
values, implying considerable learning compared to the initial flat
priors. 

In sum, from the estimates of the customer Bayesian learning models,
we conclude that considerable learning can be achieved on a period-by-period
basis with the arrival of new information and that the learning processes
may vary greatly across customers. As before, we note that these conclusions
are based on a particular learning model. The empirical relevance
of this model to predict customer decisions will be evaluated in the
remainder of this section. 

\subsection{Choice Model and Demand Arrival }

We now consider the estimation of the utility function that determines
customer choices to use AlphaShip as a service provider. This estimation
depends on i) which learning model customers use to predict quality
and ii) the shape of the shipping quality function (i.e., asymmetry,
non-linearity and risk aversion). Table \ref{Table: Model Rank} reports
a rank of goodness-of-fit for alternative specifications of the shipping
quality function (see $\mathsection$4.3.3) and learning models (see
Table \ref{Table: Benchmark Models}). Here we use all the predictors
(e.g., intercept, price) explained in $\mathsection$4.2, because
we find they are all significant for all the models and that the values
of the coefficients remain mostly unchanged across models. 

As previously mentioned, the models we compare differ on two major
dimensions: (1) the learning model and (2) the shape of the service
quality function $f\left(I_{t},\boldsymbol{\beta}^{q}\right)$. Specifically,
we compare different learning assumptions, including short memory
(i.e., benchmark model 1), independent learning (i.e., benchmark model
2), information pooling learning (i.e., benchmark model 3A) and spillover
learning (i.e., simple hierarchical model). Here we do not compare
the regression counterparts of the models because $Distance$ is not
a useful predictor based on the results in $\mathsection$5.1. As
we have discussed in $\mathsection$4.3.3, the specification of the
service quality function $f\left(I_{t},\boldsymbol{\beta}^{q}\right)$
may differ in terms of three aspects: (1) whether the effect of expected
service quality mean $\bar{Q}^{E}$ is symmetric or asymmetric; (2)
whether the effect of expected service quality $\bar{Q}^{E}$ is linear
or nonlinear; (3) whether the customers are risk averse or risk neutral,
and if the customers are risk averse, whether the customers are averse
to either one of or both of sources of uncertainty (experience variability
and mean quality uncertainty). 

We test all the specification combinations and report the likelihood
of selected models in Table \ref{Table: Model Fitting LL} in the
Appendix and provide a goodness of fit rank in Table
\begin{table}[tb]
\caption{Rank of Goodness of Fit on Alternative Specifications of the Shipping
Quality Function}
\label{Table: Model Rank}
\centering{}%
\begin{tabular}{>{\raggedright}p{3.3cm}|>{\centering}p{2cm}>{\centering}p{2cm}>{\centering}p{2cm}>{\raggedleft}p{2cm}}
\hline 
\textsf{\footnotesize{}LL/AIC} & \textsf{\footnotesize{}Benchmark Model 1 (short memory)} & \textsf{\footnotesize{}Benchmark Model 3A (information pooling)} & \textsf{\footnotesize{}Benchmark Model 2 (independent learning)} & \textsf{\footnotesize{}Simple Hierarchical Model}\tabularnewline
\hline 
\textsf{\footnotesize{}S{*}} & \textsf{\footnotesize{}16/16{*}{*}} & \textsf{\footnotesize{}15/15} & \textsf{\footnotesize{}14/14} & \textsf{\footnotesize{}13/13}\tabularnewline
\textsf{\footnotesize{}A} & \textsf{\footnotesize{}\textemdash \textendash{}} & \textsf{\footnotesize{}12/12} & \textsf{\footnotesize{}11/11} & \textsf{\footnotesize{}9/9}\tabularnewline
\textsf{\footnotesize{}A + ERA} & \textsf{\footnotesize{}\textemdash \textendash{}} & \textsf{\footnotesize{}10/10} & \textsf{\footnotesize{}8/8} & \textsf{\footnotesize{}7/7}\tabularnewline
\textsf{\footnotesize{}A + ERA + BUA} & \textsf{\footnotesize{}\textemdash \textendash{}} & \textsf{\footnotesize{}6/5} & \textsf{\footnotesize{}4/3} & \textsf{\footnotesize{}2/1}\tabularnewline
\textsf{\footnotesize{}A + ERA + BUA + Q} & \textsf{\footnotesize{}\textemdash \textendash{}} & \textsf{\footnotesize{}5/6} & \textsf{\footnotesize{}3/4} & \textsf{\footnotesize{}1/2}\tabularnewline
\hline 
\multicolumn{5}{>{\raggedright}p{11.3cm}}{\textsf{\scriptsize{}{*}: We use }\textsf{\textit{\scriptsize{}S }}\textsf{\scriptsize{}for
symmetric;}\textsf{\textit{\scriptsize{} A}}\textsf{\scriptsize{}
for }\textsf{\textit{\scriptsize{}Asymmetric}}\textsf{\scriptsize{};
}\textsf{\textit{\scriptsize{}ERA}}\textsf{\scriptsize{} for }\textsf{\textit{\scriptsize{}Experience
risk averse}}\textsf{\scriptsize{}; }\textsf{\textit{\scriptsize{}BUA}}\textsf{\scriptsize{}
for}\textsf{\textit{\scriptsize{} belief uncertainty averse}}\textsf{\scriptsize{};
}\textsf{\textit{\scriptsize{}Q}}\textsf{\scriptsize{} for }\textsf{\textit{\scriptsize{}Quadratic}}\textsf{\scriptsize{}. }{\scriptsize \par}

\textsf{\scriptsize{}{*}{*}: The rank of models' goodness-of-fit by
log-likelihood is before the slash, and that by Akaike information
criterion (AIC) is after the slash. 1 means the model fits the best
under the criterion, and 16 means the model fits the worst.}}\tabularnewline
\end{tabular}
\end{table}
\ref{Table: Model Rank}. The results of these goodness of fit comparisons
show how these measures improve with the properties of the learning
model and service quality function. Specifically, for each row of
Table \ref{Table: Model Rank}, the rank decreases from left to right,
meaning that the simple hierarchical learning model (see Table \ref{Table: Shipping Decision Results}
for the estimates) fits the data better than the independent learning
assumption under the same specification of the quality function $f\left(I_{t},\boldsymbol{\beta}^{q}\right)$
(i.e., within the same row). In addition, the fit of the independent
learning model is in turn better than that of the information pooling
learning model. Overall, these results provide evidence consistent
with information spillovers driving customer choices.

Next, if we look at each column, we find the rank by model log-likelihood
improves from top to bottom, showing that as the model becomes more
flexible, goodness-of-fit improves under the same learning model (i.e.,
the same column). This model comparison, however, does not consider
the number of parameters of each model. Accordingly, it is useful
to note that the ranks in the last two rows reverse under the criterion
of AIC, because the improvement by adding the quadratic structure
is not enough to compensate for the penalty associated with adding
the quadratic parameters. Considering both model fit and model parsimony,
we choose the asymmetric risk averse model (i.e., $A+ERA+BUA$) as
the best model, and use its coefficients under the simple hierarchical
learning model for the following discussion and the policy simulations
in section 6.
\begin{table}[tb]
\caption{Estimation Results for Shipping Choice Model}
\label{Table: Shipping Decision Results}
\centering{}\textsf{\footnotesize{}}%
\begin{tabular}{>{\raggedright}p{2.2cm}>{\centering}p{2cm}>{\centering}p{2cm}>{\centering}p{2cm}>{\centering}p{2cm}>{\centering}p{2cm}>{\centering}p{2cm}}
\hline 
\textsf{\textbf{\small{}Model}} & \textsf{\textbf{\small{}1}} & \textsf{\textbf{\small{}2}} & \textsf{\textbf{\small{}3}} & \textsf{\textbf{\small{}4}} & \textsf{\textbf{\small{}5}} & \textsf{\textbf{\small{}6}}\tabularnewline
 & \textsf{\footnotesize{}Null} & \textsf{\footnotesize{}S$^{\dagger}$} & \textsf{\footnotesize{}A} & \textsf{\footnotesize{}A + ERA} & \textsf{\footnotesize{}A + ERA + BUA} & \textsf{\footnotesize{}A + ERA + BUA + Q}\tabularnewline
\hline 
\multicolumn{7}{l}{\textsf{\textit{\small{}Shipping Choice Parameters}}}\tabularnewline
\textsf{\textit{\footnotesize{}Intercept}} & \textsf{\footnotesize{}-1.74e-1{*}{*}{*}}{\footnotesize \par}

\textsf{\footnotesize{}(2.74e-2)} & \textsf{\footnotesize{}-1.76e-1{*}{*}{*}}{\footnotesize \par}

\textsf{\footnotesize{}(2.74e-2)} & \textsf{\footnotesize{}-1.73e-1{*}{*}{*}}{\footnotesize \par}

\textsf{\footnotesize{}(2.74e-2)} & \textsf{\footnotesize{}-1.67e-1{*}{*}{*}}{\footnotesize \par}

\textsf{\footnotesize{}(2.70e-2)} & \textsf{\footnotesize{}-1.31e-1{*}{*}{*}}{\footnotesize \par}

\textsf{\footnotesize{}(2.58e-2)} & \textsf{\footnotesize{}-1.26e-2{*}{*}{*}}{\footnotesize \par}

\textsf{\footnotesize{}(2.35e-2)}\tabularnewline
\textsf{\textit{\footnotesize{}$\Omega$(Intercept)}} & \textsf{\footnotesize{}6.05e-2{*}{*}{*}}{\footnotesize \par}

\textsf{\footnotesize{}(2.65e-2)} & \textsf{\footnotesize{}6.06e-2{*}{*}{*}}{\footnotesize \par}

\textsf{\footnotesize{}(2.62e-2)} & \textsf{\footnotesize{}6.05e-2{*}{*}{*}}{\footnotesize \par}

\textsf{\footnotesize{}(2.65e-2)} & \textsf{\footnotesize{}6.05e-2{*}{*}{*}}{\footnotesize \par}

\textsf{\footnotesize{}(2.64e-2)} & \textsf{\footnotesize{}5.89e-2{*}{*}{*}}{\footnotesize \par}

\textsf{\footnotesize{}(2.51e-2)} & \textsf{\footnotesize{}5.88e-2{*}{*}{*}}{\footnotesize \par}

\textsf{\footnotesize{}(2.51e-2)}\tabularnewline
\textsf{\textit{\footnotesize{}Price}} & \textsf{\footnotesize{}-1.44e-1{*}{*}{*}}{\footnotesize \par}

\textsf{\footnotesize{}(1.06e-2)} & \textsf{\footnotesize{}-1.44e-1{*}{*}{*}}{\footnotesize \par}

\textsf{\footnotesize{}(1.06e-2)} & \textsf{\footnotesize{}-1.46e-1{*}{*}{*}}{\footnotesize \par}

\textsf{\footnotesize{}(1.07e-2)} & \textsf{\footnotesize{}-1.44e-1{*}{*}{*}}{\footnotesize \par}

\textsf{\footnotesize{}(1.07e-2)} & \textsf{\footnotesize{}-1.42e-1{*}{*}{*}}{\footnotesize \par}

\textsf{\footnotesize{}(1.08e-2)} & \textsf{\footnotesize{}-1.42e-1{*}{*}{*}}{\footnotesize \par}

\textsf{\footnotesize{}(1.07e-2)}\tabularnewline
\textsf{\textit{\footnotesize{}Weight }} & \textsf{\footnotesize{}3.64e-1{*}{*}{*}}{\footnotesize \par}

\textsf{\footnotesize{}(1.54e-2)} & \textsf{\footnotesize{}3.64e-1{*}{*}{*}}{\footnotesize \par}

\textsf{\footnotesize{}(1.54e-2)} & \textsf{\footnotesize{}3.64e-1{*}{*}{*}}{\footnotesize \par}

\textsf{\footnotesize{}(1.54e-2)} & \textsf{\footnotesize{}3.63e-1{*}{*}{*}}{\footnotesize \par}

\textsf{\footnotesize{}(1.54e-2)} & \textsf{\footnotesize{}3.62e-1{*}{*}{*}}{\footnotesize \par}

\textsf{\footnotesize{}(1.54e-2)} & \textsf{\footnotesize{}3.62e-1{*}{*}{*}}{\footnotesize \par}

\textsf{\footnotesize{}(1.54e-2)}\tabularnewline
\textsf{\textit{\footnotesize{}SecondHalfWeek}} & \textsf{\footnotesize{}-5.54e-2{*}{*}{*}}{\footnotesize \par}

\textsf{\footnotesize{}(8.77e-3)} & \textsf{\footnotesize{}-5.54e-2{*}{*}{*}}{\footnotesize \par}

\textsf{\footnotesize{}(8.77e-3)} & \textsf{\footnotesize{}-5.52e-2{*}{*}{*}}{\footnotesize \par}

\textsf{\footnotesize{}(8.76e-3)} & \textsf{\footnotesize{}-5.52e-2{*}{*}{*}}{\footnotesize \par}

\textsf{\footnotesize{}(8.79e-3)} & \textsf{\footnotesize{}-5.53e-2{*}{*}{*}}{\footnotesize \par}

\textsf{\footnotesize{}(8.54e-3)} & \textsf{\footnotesize{}-5.33e-2{*}{*}{*}}{\footnotesize \par}

\textsf{\footnotesize{}(8.56e-3)}\tabularnewline
\textsf{\textit{\footnotesize{}$\mu_{jt}$}} & \textsf{\footnotesize{}\textemdash \textendash{}} & \textsf{\footnotesize{}-5.01e-2{*}{*}{*}}{\footnotesize \par}

\textsf{\footnotesize{}(1.02e-2)} & \textsf{\footnotesize{}\textemdash \textendash{}} & \textsf{\footnotesize{}\textemdash \textendash{}} & \textsf{\footnotesize{}\textemdash \textendash{}} & \textsf{\footnotesize{}\textemdash \textendash{}}\tabularnewline
\textsf{\textit{\footnotesize{}$\left[\mu_{jt}\right]^{+}$}} & \textsf{\footnotesize{}\textemdash \textendash{}} & \textsf{\footnotesize{}\textemdash \textendash{}} & \textsf{\footnotesize{}-7.98e-2{*}{*}{*}}{\footnotesize \par}

\textsf{\footnotesize{}(3.67e-2)} & \textsf{\footnotesize{}-8.03e-2{*}{*}{*}}{\footnotesize \par}

\textsf{\footnotesize{}(3.67e-2)} & \textsf{\footnotesize{}-8.71e-2{*}{*}{*}}{\footnotesize \par}

\textsf{\footnotesize{}(3.88e-2)} & \textsf{\footnotesize{}-9.13e-1{*}{*}{*}}{\footnotesize \par}

\textsf{\footnotesize{}(4.12e-2)}\tabularnewline
\textsf{\textit{\footnotesize{}$\left[\mu_{jt}\right]^{-}$}} & \textsf{\footnotesize{}\textemdash \textendash{}} & \textsf{\footnotesize{}\textemdash \textendash{}} & \textsf{\footnotesize{}-4.69e-2{*}{*}{*}}{\footnotesize \par}

\textsf{\footnotesize{}(2.11e-2)} & \textsf{\footnotesize{}-4.65e-2{*}{*}{*}}{\footnotesize \par}

\textsf{\footnotesize{}(2.10e-2)} & \textsf{\footnotesize{}-4.98e-2{*}{*}{*}}{\footnotesize \par}

\textsf{\footnotesize{}(2.13e-2)} & \textsf{\footnotesize{}-5.22e-2{*}{*}}{\footnotesize \par}

\textsf{\footnotesize{}(2.19e-2)}\tabularnewline
\textsf{\textit{\footnotesize{}$\sigma^{2}$}} & \textsf{\footnotesize{}\textemdash \textendash{}} & \textsf{\footnotesize{}\textemdash \textendash{}} & \textsf{\footnotesize{}\textemdash \textendash{}} & \textsf{\footnotesize{}-3.73e-2{*}{*}{*}}{\footnotesize \par}

\textsf{\footnotesize{}(1.02e-2)} & \textsf{\footnotesize{}-3.75e-2{*}{*}{*}}{\footnotesize \par}

\textsf{\footnotesize{}(1.02e-2)} & \textsf{\footnotesize{}-3.76e-2{*}{*}{*}}{\footnotesize \par}

\textsf{\footnotesize{}(1.02e-2)}\tabularnewline
\textsf{\footnotesize{}$\Omega\left(\sigma^{2}\right)$} & \textsf{\footnotesize{}\textemdash \textendash{}} & \textsf{\footnotesize{}\textemdash \textendash{}} & \textsf{\footnotesize{}\textemdash \textendash{}} & \textsf{\footnotesize{}1.38e-2{*}{*}{*}}{\footnotesize \par}

\textsf{\footnotesize{}(6.20e-3)} & \textsf{\footnotesize{}1.39e-2{*}{*}{*}}{\footnotesize \par}

\textsf{\footnotesize{}(6.20e-3)} & \textsf{\footnotesize{}1.39e-2{*}{*}{*}}{\footnotesize \par}

\textsf{\footnotesize{}(6.22e-3)}\tabularnewline
\textsf{\textit{\footnotesize{}$Var\left(\mu_{jt}\right)$$^{\divideontimes}$}} & \textsf{\footnotesize{}\textemdash \textendash{}} & \textsf{\footnotesize{}\textemdash \textendash{}} & \textsf{\footnotesize{}\textemdash \textendash{}} & \textsf{\footnotesize{}\textemdash \textendash{}} & \textsf{\footnotesize{}-6.76e-2{*}{*}{*}}{\footnotesize \par}

\textsf{\footnotesize{}(1.91e-2)} & \textsf{\footnotesize{}-6.75e-2{*}{*}{*}}{\footnotesize \par}

\textsf{\footnotesize{}(1.91e-2)}\tabularnewline
\textsf{\footnotesize{}$\Omega\left(Var\left(\mu_{jt}\right)\right)$} & \textsf{\footnotesize{}\textemdash \textendash{}} & \textsf{\footnotesize{}\textemdash \textendash{}} & \textsf{\footnotesize{}\textemdash \textendash{}} & \textsf{\footnotesize{}\textemdash \textendash{}} & \textsf{\footnotesize{}2.37e-2{*}{*}{*}}{\footnotesize \par}

\textsf{\footnotesize{}(8.43e-3)} & \textsf{\footnotesize{}2.36e-2{*}{*}{*}}{\footnotesize \par}

\textsf{\footnotesize{}(8.41e-3)}\tabularnewline
\textsf{\textit{\footnotesize{}$\left(\left[\mu_{jt}\right]^{+}\right)^{2}$}} & \textsf{\footnotesize{}\textemdash \textendash{}} & \textsf{\footnotesize{}\textemdash \textendash{}} & \textsf{\footnotesize{}\textemdash \textendash{}} & \textsf{\footnotesize{}\textemdash \textendash{}} & \textsf{\footnotesize{}\textemdash \textendash{}} & \textsf{\footnotesize{}2.60e-3}{\footnotesize \par}

\textsf{\footnotesize{}(1.12e-2)}\tabularnewline
\textsf{\textit{\footnotesize{}$\left(\left[\mu_{jt}\right]^{-}\right)^{2}$}} & \textsf{\footnotesize{}\textemdash \textendash{}} & \textsf{\footnotesize{}\textemdash \textendash{}} & \textsf{\footnotesize{}\textemdash \textendash{}} & \textsf{\footnotesize{}\textemdash \textendash{}} & \textsf{\footnotesize{}\textemdash \textendash{}} & \textsf{\footnotesize{}2.30e-3}{\footnotesize \par}

\textsf{\footnotesize{}(7.40e-3)}\tabularnewline
\hline 
\multicolumn{7}{l}{\textsf{\textit{\small{}Demand Arrival Parameters (mean and 95\% quantile
interval)}}}\tabularnewline
\textsf{\textbf{\footnotesize{}$\max_{j\in\Upsilon_{i}}\left(m_{ij}\right)$}} & \textsf{\footnotesize{}0.539}{\footnotesize \par}

\textsf{\footnotesize{}(0.317, 0.734)} & \textsf{\footnotesize{}0.538}{\footnotesize \par}

\textsf{\footnotesize{}(0.317, 0.735)} & \textsf{\footnotesize{}0.540}{\footnotesize \par}

\textsf{\footnotesize{}(0.319, 0.748)} & \textsf{\footnotesize{}0.542}{\footnotesize \par}

\textsf{\footnotesize{}(0.319, 0.736)} & \textsf{\footnotesize{}0.541}{\footnotesize \par}

\textsf{\footnotesize{}(0.319, 0.736)} & \textsf{\footnotesize{}0.541}{\footnotesize \par}

\textsf{\footnotesize{}(0.319, 0.739)}\tabularnewline
\textsf{\textbf{\footnotesize{}$\min_{j\in\Upsilon_{i}}\left(m_{ij}\right)$}} & \textsf{\footnotesize{}0.180}{\footnotesize \par}

\textsf{\footnotesize{}(0.024, 0.440)} & \textsf{\footnotesize{}0.182}{\footnotesize \par}

\textsf{\footnotesize{}(0.024, 0.443)} & \textsf{\footnotesize{}0.183}{\footnotesize \par}

\textsf{\footnotesize{}(0.025, 0.443)} & \textsf{\footnotesize{}0.184}{\footnotesize \par}

\textsf{\footnotesize{}(0.027, 0.443)} & \textsf{\footnotesize{}0.184}{\footnotesize \par}

\textsf{\footnotesize{}(0.026, 0.445)} & \textsf{\footnotesize{}0.185}{\footnotesize \par}

\textsf{\footnotesize{}(0.025, 0.446)}\tabularnewline
\hline 
\textsf{\footnotesize{}$LL$} & \textsf{\footnotesize{}-53648.6} & \textsf{\footnotesize{}-53623.6} & \textsf{\footnotesize{}-53582.1} & \textsf{\footnotesize{}-53569.7} & \textsf{\footnotesize{}-53544.2} & \textsf{\footnotesize{}-53544.0}\tabularnewline
\hline 
\multicolumn{7}{>{\raggedright}p{13.2cm}}{\textsf{\footnotesize{}$^{\dagger}$: We use }\textsf{\textit{\footnotesize{}S
}}\textsf{\footnotesize{}for symmetric;}\textsf{\textit{\footnotesize{}
A}}\textsf{\footnotesize{} for }\textsf{\textit{\footnotesize{}asymmetric}}\textsf{\footnotesize{};
}\textsf{\textit{\footnotesize{}ERA}}\textsf{\footnotesize{} for }\textsf{\textit{\footnotesize{}experience
risk averse}}\textsf{\footnotesize{}; }\textsf{\textit{\footnotesize{}BUA}}\textsf{\footnotesize{}
for}\textsf{\textit{\footnotesize{} belief uncertainty averse}}\textsf{\footnotesize{};
}\textsf{\textit{\footnotesize{}Q}}\textsf{\footnotesize{} for }\textsf{\textit{\footnotesize{}quadratic}}\textsf{\footnotesize{}. }{\footnotesize \par}

\textsf{\footnotesize{}$^{\divideontimes}$: Var($\mu_{jt}$)$=\xi^{2}+\sigma_{\mu}^{2}$
under the simple hierarchical model.}{\footnotesize \par}

\textsf{\footnotesize{}{*}p<0.1, {*}{*}p<0.05, {*}{*}{*}p<0.01}{\footnotesize \par}

\textsf{\footnotesize{}Note. Most of the predictors are normalized
to have zero mean and approximate 1 standard deviation. Specifically,
$\sigma^{2}$ and $Var\left(\mu_{jt}\right)$ are scaled by 10; $\mu_{jt}$,
$\left[\mu_{jt}\right]^{+}$ and $\left[\mu_{jt}\right]^{-}$ are
scaled by 2; Price is scaled by 5000; Weight is scaled by 3000; CR
is scaled by 6.}}\tabularnewline
\end{tabular}{\footnotesize \par}
\end{table}

\subsubsection{Results for Shipping Choice and Demand Arrival Model}

The coefficients estimated for alternative models under different
learning models are very similar (i.e., same signs and similar magnitudes),
so we only provide the results under the simple hierarchical learning
specification (see Table \ref{Table: Shipping Decision Results}).
In this table the variance of the random coefficients for a particular
predictor is indicated with the $\Omega$ symbol (e.g., $\Omega(Intercept)$
denotes the variance of the random intercepts across customers). These
variances measure the extent of heterogeneity for the effect of each
utility predictor across customers. 

As expected, the log-likelihood of customer choices increases from
left to right as the utility model becomes more flexible in terms
of the service quality function. We focus on the asymmetric risk averse
model (i.e., $A+ERA+BUA$) which provides the best balance between
fit and parsimony (model 5). In particular, allowing for an asymmetric
effect of the predicted mean quality, we find that customers are more
sensitivity to tardiness than earliness. Note that the coefficients
for both tardiness (\textsf{\textit{\footnotesize{}$\left[\mu_{jt}\right]^{+}$}})
and earliness (\textsf{\textit{\footnotesize{}$\left[\mu_{jt}\right]^{-}$}})
are both negative, however the magnitude of the first coefficient
is estimated to be greater. This implies customers exhibit a negative
response to tardiness, a positive response to earliness and that negative
responses to tardiness are approximately 1.7 times greater than positive
responses to earliness. We also tried adding random coefficients for
tardiness (\textsf{\textit{\footnotesize{}$\left[\mu_{jt}\right]^{+}$}})
and earliness (\textsf{\textit{\footnotesize{}$\left[\mu_{jt}\right]^{-}$}})
to allow for heterogeneity across customers on these two dimensions,
however the log-likelihood increase is very small (i.e., around 3
points).

In terms of the impact of uncertainty, we find that customers are
averse to experience variability ($\sigma^{2}$), but even more so
to belief uncertainty ($Var(\mu_{jt})$). This finding again confirms
the importance of taking into account both sources of uncertainty.
Our results show that even though customers are averse to both risks,
the impact of belief uncertainty is almost twice as large as that
of the experience variability. We also note that by comparing models
5 and 6 we find that adding quadratic terms to the service quality
function yields almost no improvement in the log-likelihood, and this
is consistent with the quadratic coefficients being non-significant. 

Finally, recall that the model also includes demand arrival parameters
$\bar{\boldsymbol{m}}_{i}=\left\{ \bar{m}_{ij},\forall j\in\Upsilon_{i}\right\} $,
which allow some routes for a given customer to be more frequently
demanded than others (see equation \ref{eq: Demand Arrival}). We
obtained demand arrival parameters $\bar{\boldsymbol{m}}_{i}$ for
each customer $i$, rendering $3000+$ parameters in total. Given
that customers have different numbers of routes, we only provide the
population summary of each customer's most and least frequent route.
On average, the top route for a given customer is roughly three times
more popular than the least demanded route. 

\subsection{Robustness to Choice Set (Route) Definition}

In addition to using routes as choice alternatives for each customer,
we have also used a more aggregated definition based on destination
countries. Given that the customers in our data have fixed physical
locations and usually rely on only a single origin airport, different
routes of a customer are actually differentiated by different destination
airports. To test the model robustness to different levels of product
aggregation, we aggregated the destination airports into destination
countries and consistent with our spillover formulation allowed customers'
experiences shipping to one country to affect their expectations about
service quality when shipping to other countries. After re-estimating
all models we obtain virtually the same results: the information spillover
model not only outperforms other learning models at predicting shipping
service qualities, but also in terms of predicting customers' purchase
decisions. These results again provide evidence consistent with information
spillovers driving customer choices. 

\section{Policy Simulations and Managerial Implications\label{sec: Applications}}

The results of the previous section suggest that customer choices
are consistent with the use of quality information from other products
to form beliefs about a particular product. In addition, customers
are sensitive to not only the predicted mean service quality but also
to quality uncertainty. In this section, we discuss managerial insights
directly related to these findings. More specifically, we study the
consequences of unsatisfactory levels of service quality. When service
quality decreases on certain routes (e.g., due to higher average transport
delay or larger experience variability), our model can be used to
predict the corresponding impact on customer behavior. 

For simplicity, we consider customers with routes $A$ and $B$ and
whose demand arrival on the two routes follows a multinomial $\left(0.5,0.5\right)$
distribution. Transport delays on these two route, $Q_{A}$ and $Q_{B}$
respectively, follow the same normal distribution, $Q_{A}\sim N\left(0,5^{2}\right)$
and $Q_{B}\sim N\left(0,5^{2}\right)$. Customers decide whether to
ship based on the utility of shipping compared to that of the outside
option and then update their quality beliefs if new information is
acquired (i.e., if they used AlphaShip services).\footnote{When calculating utilities we set other exogenous variables (e.g.,
price) to their empirical means.} In order to examine the effect of an increase in the average transport
delay, we change the transport delay on route $A$ to $Q_{A}\sim N\left(5,5^{2}\right)$
starting from period 20. 
\begin{figure}[tb]
\begin{centering}
\begin{minipage}[t]{0.33\columnwidth}%
\includegraphics[width=1\columnwidth]{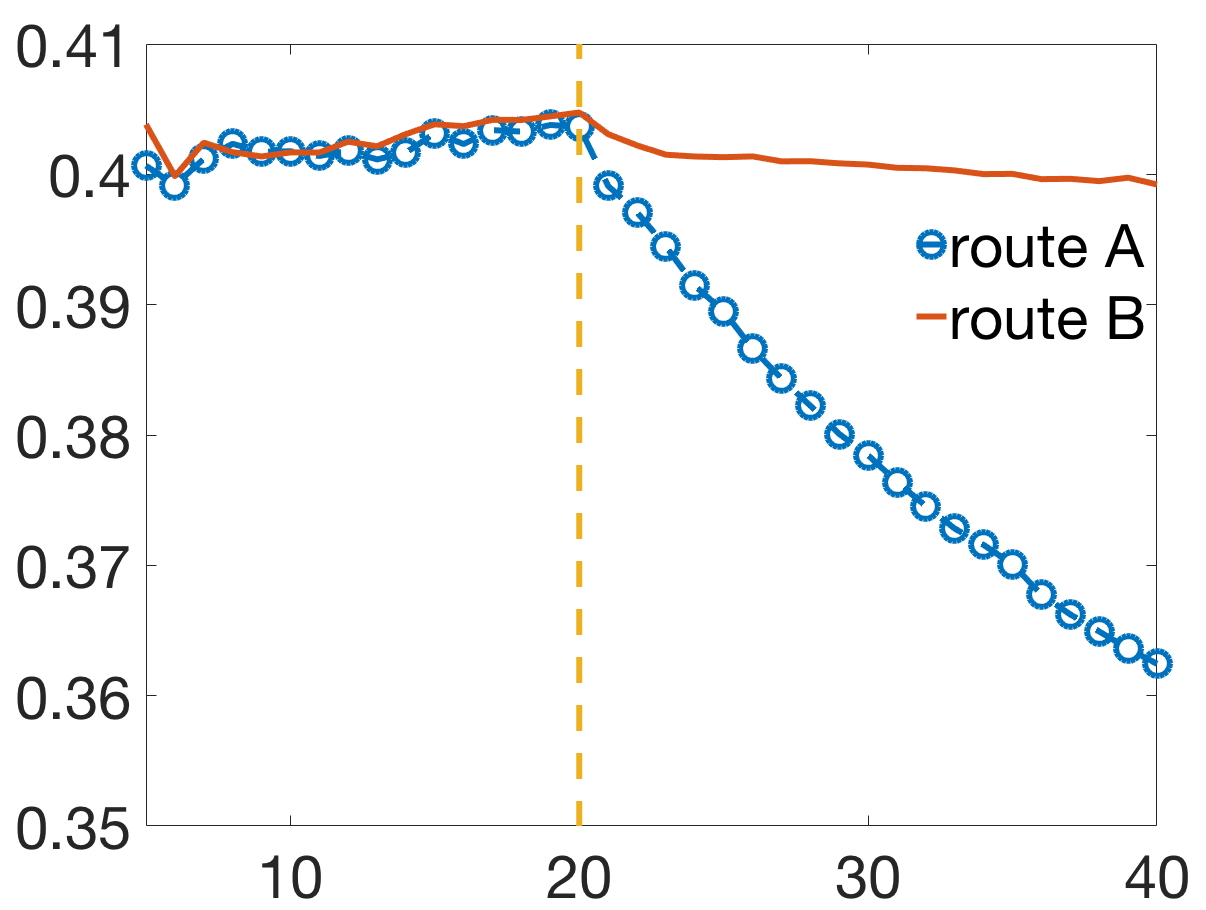}%
\end{minipage}%
\begin{minipage}[t]{0.32\columnwidth}%
\includegraphics[width=1\columnwidth]{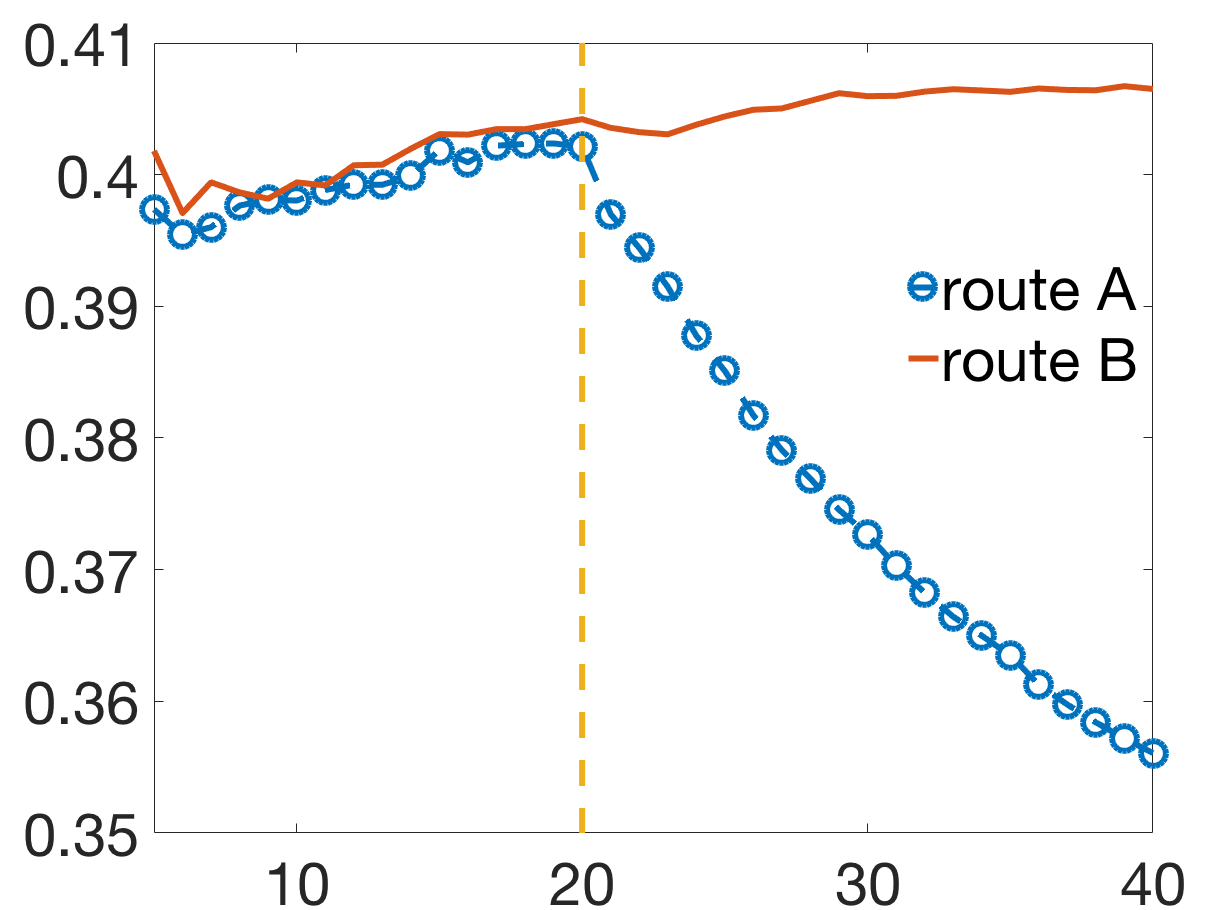}%
\end{minipage}%
\begin{minipage}[t]{0.33\columnwidth}%
\includegraphics[width=1\columnwidth]{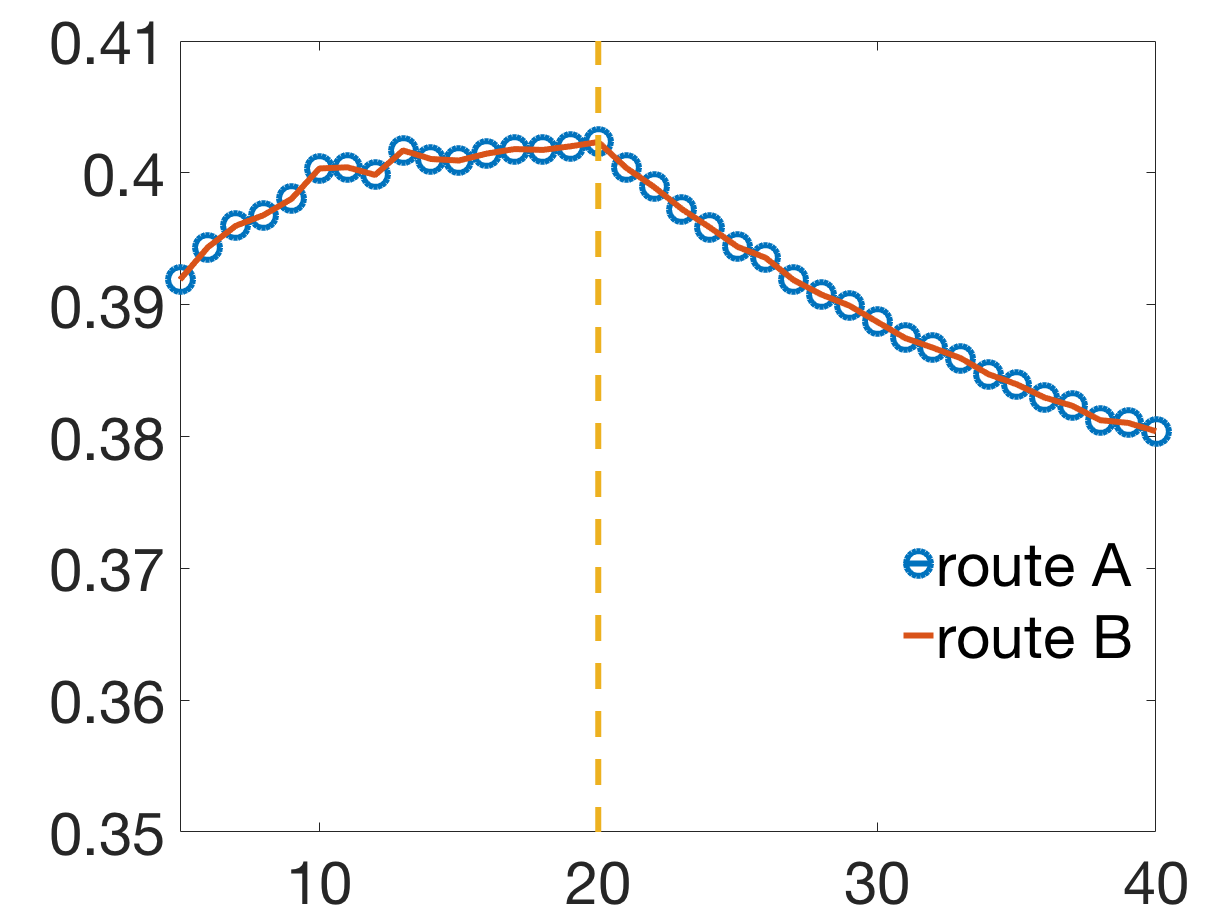}%
\end{minipage}
\par\end{centering}
\centering{}\caption{Purchase probability on route A and B assuming different learning
rules. Left: spillover learning; Center: independent learning; Right:
information pooling learning.}
\label{Figure: Counterfactual}
\end{figure}
The left panel in Figure \ref{Figure: Counterfactual} presents the
average purchase probability on the two routes over 40 periods.\footnote{The average purchase probability is the mean of purchase probability
of 200 simulated customers, and the customers' quality sensitivities
are simulated from the distribution of the estimated random coefficients.} As we can see from the left panel (spillover learning), before the
quality change on route $A$ in period 20 (marked by the vertical
line in Figure \ref{Figure: Counterfactual}), the purchase probabilities
on routes $A$ and $B$ are very similar and both increase over time,
where the increase in purchase probability is caused by the decrease
in learning uncertainty as customers collect more information. However,
after the average transport delay on route $A$ increases from 0 to
5 in period 20, the average purchase probability on route $A$ drops
rapidly due to the deteriorating transport service quality. Interestingly,
the purchase probability on route $B$ stops increasing and also drops
after period 20. Given that the service quality on route $B$ never
changes, the drop in purchase probability is exclusively caused by
the spillover of customers' experiences from deteriorating service
quality on route $A$. If we use the average empirical price of \$2,288
as the revenue obtained from a shipment on route $A$ and $B$, then
when reaching the last period the increase of transport delay on route
$A$ causes a direct expected loss of \$91.5 per customer on route
$A$ and an indirect expected loss of \$7.2 per customer on route
$B$. 

This policy simulation also illustrates the consequences of using
an alternative learning model, such as independent learning or information
pooling. Specifically, the center and right panel of Figure \ref{Figure: Counterfactual}
show the changes in purchase probability on routes $A$ and $B$ under
independent learning and information pooling, respectively. As we
can see, under independent learning, the quality deterioration on
route $A$ won't be expected to affect the demand on route $B$. In
fact, as customers acquire more information, the purchase probability
on route $B$ increases over time. In contrast, under information
pooling, experiences from routes $A$ and $B$ are equally informative
about either route; thus the purchase probability for these two routes
is identical. Moreover, the low service quality experiences on route
$A$ are compensated by the good experiences on route $B$. Hence,
the decrease of purchase probability on route $A$ is smaller than
that under information spillover learning or independent learning.
As before, we use an average price of \$2,288 for the revenue obtained
from a shipment on route $A$ and $B$. If instead of accounting for
spillovers, customers were assumed to be independent learners, the
increase of transport delay on route $A$ would be anticipated to
cause only a direct expected loss of \$107.2 on route $A$, while
no indirect loss on route $B$ would be expected. If instead customers
were assumed to adopt information pooling as a learning rule, the
expected revenue loss on route $A$ and $B$ would be identical and
equal to \$50.3. 

The results of this policy simulation illustrates that when making
decisions about whether to improve service quality and particularly
how much budget to allocate to quality improvements, it is important
to consider and account for information spillovers. This is because
the financial consequences of service improvements for each product
may be strongly affected by the learning processes that customers
use when relying on previous experiences across multiple services
to form quality beliefs about future service encounters. Another counterfactual
experiment comparing the effects of deteriorating quality mean and
variance is provided in Appendix \ref{sec:Supplementary-Application}. 

\section{Conclusions, Limitations and Future Direction\label{sec: Conclusions}}

In this paper, we used a one-year shipping and sales historical data
of individual customers of a world-leading third-party logistics company
to study how customers learn about shipping service quality from their
past experiences, especially from experiences of similar yet not identical
service encounters. This learning behavior is referred to as ``correlated
learning'' or ``spillover learning'' in the learning literature.
Most of the previous studies in this literature do not directly observe
quality and hence rely on imputing quality or on surveys for gathering
subjective measures of product or service quality. Our dataset instead
contains objective service quality measures. This advantage allows
us to flexibly estimate customers' learning and its impact on shipping
choices. This further enables us to estimate learning models that
describe beliefs for more than 2,000 products (i.e., routes). 

An important contribution of this paper is methodological. In contrast
to the approaches that have been used in the correlated learning literature,
we use a Bayesian hierarchical methodology to model the learning process.
This model provides a natural way to represent the information borrowing
process and can be easily applied to problems with more complex correlation
structures without adding too many parameters. 

We derive insights by comparing the Bayesian hierarchical model with
three benchmark learning models: (1) a short-memory learning model;
(2) a model of independent learning; and (3) a model of information
pooling. Our results show that the hierarchical model not only outperforms
alternative models in forecasting transport delays, but also achieves
the best goodness of fit when predicting customer purchases. This
finding confirms that customer choices are best predicted by a learning
model that allows experiences about one product to influence beliefs
about other products. In addition, we find asymmetries in how customers
react to earliness and tardiness \textendash{} negative responses
to delays are greater than positive responses to the early arrival
of a shipment. We further obtain evidence of customers being risk
averse and we are able to separate the effect of experience variability
from belief uncertainty, with the latter having a greater impact on
customer choices. Finally, using counterfactual experiments, we illustrate
how quality deterioration on one product affects not only the revenues
of that product, but also the revenues of other products through learning
spillovers. Therefore, decisions about operational quality improvements
should account for both the direct and indirect effects of these improvements
on customer behavior. 

Our study can be extended in several ways in future research. One
possible extension is to measure how observable customer characteristics
\textemdash{} such as company demographics \textemdash{} are related
to the customer learning process. This would be useful, for example,
to develop targeted service improvement strategies. Competition could
also be an important aspect to consider; this would probably require
data from multiple leading companies in the market to study how the
market structure mediates the effect of service quality on customer
purchases. On a final note, this study showcases the importance of
bringing advanced methodologies in statistics, economics and marketing
into the the fields of operations management. We hope that this work
stimulates further research on the interface between these academic
disciplines.


\bibliographystyle{ormsv080}
\bibliography{KN_MSS}

\clearpage{}

\begin{APPENDICES} 

\section{Data Selection and Missing Data Interpolation}

This section explains the data manipulation steps we used to generate
the data set for model estimation.

\subsection{Data Selection\label{subsec:Data-Selection}}

We focus on customers with at least 5 shipments during January to
March (to construct the pre-estimation sample that is explained in
$\mathsection$4.3.1) and at least 15 shipments in the last 9 months
in 2013 (to construct the data for model estimation). Among the customers
that have shipped no less than 5 times in the year, only 12\% of them
have only shipped on one route. Because we are interested in customer's
learning spillover, specifically across multiple routes, and to limit
the computational complexity, we limit our focus to customers with
shipping experiences on $2-10$ routes. For the same reason, we only
consider customers whose most frequent route accounts for no more
than 70\% of his total shipments during the year. In addition, we
also exclude customers with too many shipments (exceeding 100 shipments
in our case), because these customers may have deep long-term business
partnership with AlphaShip and share integrated information system
that make them very reluctant to change service suppliers in a short
period (such as the one year time in our data) even after experiencing
poor services. We further exclude data with obvious errors (e.g.,
shipments with arrival times earlier than their departure times). 

\subsection{Missing Data Interpolation\label{subsec:Missing-Data-Interpolation}}

Our data, the purchase data, records the shipping information of each
shipment, as described in $\mathsection$3. We only observe these
information of the purchased shipping services, but in order to estimate
the choice model (Equation (\ref{eq: RUM})), we need to reconstruct
the data series to include the control variables $X_{ijt}$ and $Price_{ijt}$
for the periods when there is no purchase and for all routes of each
customer. The method to reconstruct price data is described in $\mathsection$3.3.
For the cargo-related information in $X_{ijt}$ \textemdash{} cargo
volume and pieces \textendash{} for simplicity we use the customer-route
level average to interpolate the missing data. 

\section{Supplementary Material for Customers' Bayesian Learning Model}

This section includes the supplementary material for the estimation
of customers' Bayesian learning model. 

\subsection{Pre-estimation for Customer-level Priors\label{subsec: Priors}}

\begin{table}[tb]
\caption{Pre-estimation Priors and Model Priors}
\label{Table: Pre-estimation Priors}
\centering{}%
\begin{tabular}{>{\raggedright}m{0.6cm}>{\centering}m{1cm}>{\raggedleft}m{0.1cm}>{\centering}m{1.2cm}>{\raggedleft}m{3cm}}
\hline 
\multirow{2}{0.6cm}{} & \textsf{\footnotesize{}$t=1$} &  & \multicolumn{2}{c}{\textsf{\footnotesize{}$t=25$}}\tabularnewline
\cline{2-2} \cline{4-5} 
 &  &  & \textsf{\footnotesize{}Mean} & \textsf{\footnotesize{}Quantile (5\%, 95\%)}\tabularnewline
\hline 
\textsf{\textbf{\footnotesize{}$\alpha_{\sigma}$}} & \textsf{\footnotesize{}1.05} &  & \textsf{\footnotesize{}5.39} & \textsf{\footnotesize{}(3.55, 8.55)}\tabularnewline
\textsf{\textbf{\footnotesize{}$\delta_{\sigma}$}} & \textsf{\footnotesize{}10} &  & \textsf{\footnotesize{}40.95} & \textsf{\footnotesize{}(14.13, 118.03)}\tabularnewline
\textsf{\textbf{\footnotesize{}$\mu_{0}$}} & \textsf{\footnotesize{}0} &  & \textsf{\footnotesize{}-0.77} & \textsf{\footnotesize{}(-3.26, 1.77)}\tabularnewline
\textsf{\textbf{\footnotesize{}$\sigma_{\mu}$}} & \textsf{\footnotesize{}30} &  & \textsf{\footnotesize{}1.15} & \textsf{\footnotesize{}(0.36, 2.17)}\tabularnewline
\textsf{\textbf{\footnotesize{}$\alpha_{\xi}$}} & \textsf{\footnotesize{}1.05} &  & \textsf{\footnotesize{}2.37} & \textsf{\footnotesize{}(2.05, 3.05)}\tabularnewline
\textsf{\textbf{\footnotesize{}$\delta_{\xi}$}} & \textsf{\footnotesize{}3} &  & \textsf{\footnotesize{}9.51} & \textsf{\footnotesize{}(5.65, 16.13)}\tabularnewline
\hline 
\end{tabular}
\end{table}
At the beginning of period 1, all customers have the same quality
priors (see column $t=1$ in Table \ref{Table: Pre-estimation Priors}).
After calibrating the individual learning process for each customer
using the pre-estimation sample (i.e., the first 24 periods data),
we obtain quality priors for each customer at the beginning of period
25. The column $t=25$ of Table \ref{Table: Pre-estimation Priors}
provides summary statistics of these customer-level quality priors. 

\subsection{Advantages of Gibbs Sampling\label{subsec:Advantages-of-Gibbs}}

In Bayesian statistics, the posterior distribution is often not available
analytically, typically because the posterior distribution requires
the computation of a normalizing constant which is typically not available
in closed form. For this reason, posterior calculations usually rely
on Markov chain Monte Carlo (MCMC) sampling. The basic idea in MCMC
sampling is to construct a Markov chain having a stationary distribution
corresponding to the joint posterior distribution of the model parameters,
with this done in a manner that avoids ever having to calculate the
intractable constant.

In order for the Markov chain to have the appropriate behavior, the
Markov transition kernel needs to be carefully chosen, with usual
choices corresponding to either Metropolis\textendash Hastings (MH)
or Gibbs sampling. MH can involve a substantial degree of tuning for
models with many parameters, while Gibbs sampling avoids tuning by
sampling sequentially from the full-conditional posterior distributions
of subsets of parameters given current values of the other parameters.
Gibbs sampling relies on a property known as conditional conjugacy.
Focusing on a subset of the model parameters and conditioning on the
other parameters, the prior probability distribution is conditionally
conjugate if the conditional posterior distribution takes the same
form as the prior. The specific choices of our model form and prior
distributions are motivated by retaining conditional conjugacy (see
Equation (\ref{eq: True Distribution}), (\ref{eq: Simple Learning}),
(\ref{eq: Simple Hierarchy}) and (\ref{eq: Simple Prior})). 

\subsection{Gibbs Sampling of the Regression Hierarchical Model\label{subsec:Gibbs-Sampling-of}}

Gibbs sampling of the full proposed model in $\mathsection$4.3.3
can be conducted by using the following full conditional distributions
of the model parameters, and the sampling process is iterated multiple
times until convergency (we use ``$\cdots$'' to indicate all the
other parameters and data):
\begin{eqnarray}
\theta_{j}\mid\cdots & \sim & N\left(\frac{n_{j}\xi^{2}\bar{G}_{j}+\sigma^{2}\mu}{n_{j}\xi^{2}+\sigma^{2}},\;\frac{\xi^{2}\sigma^{2}}{n_{j}\xi^{2}+\sigma^{2}}\right),\nonumber \\
\mu^{d}\mid\cdots & \sim & N\left(\frac{\sigma_{\mu^{d}}^{2}G^{d}+\sigma^{2}\mu_{0}^{d}}{n^{d}\sigma_{\mu^{d}}^{2}+\sigma^{2}},\;\frac{\sigma_{\mu^{d}}^{2}\sigma^{2}}{n^{d}\sigma_{\mu^{d}}^{2}+\sigma^{2}}\right),\nonumber \\
\sigma^{2}\mid\cdots & \sim & IG\left(\alpha_{\sigma}+\frac{1}{2}\sum_{j\in\Upsilon_{i}}n_{j},\;\delta_{\sigma}+\frac{1}{2}\sum_{j\in\Upsilon_{i}}\sum_{p=1}^{t}y_{jp}^{*}\left(Q_{jp}-\theta_{j}-\mu^{d}\cdot Distance_{j}\right)^{2}\right),\nonumber \\
\mu\mid\cdots & \sim & N\left(\frac{J\sigma_{\mu}^{2}\bar{\theta}+\xi^{2}\mu_{0}}{J\sigma_{\mu}^{2}+\xi^{2}},\;\frac{\sigma_{\mu}^{2}\xi^{2}}{J\sigma_{\mu}^{2}+\xi^{2}}\right),\nonumber \\
\xi^{2}\mid\cdots & \sim & IG\left(\alpha_{\xi}+\frac{J}{2},\;\delta_{\xi}+\frac{1}{2}\sum_{j\in\Upsilon_{i}}\left(\mu_{j}-\mu\right)^{2}\right).
\end{eqnarray}
where $y_{ijt}^{*}$ is an indicator that is set to 1 if customer
$i$ has a shipment on route $j$ delivered during period $t$, and
0 otherwise. Specifically, $J=\left|\Upsilon_{i}\right|$, $\bar{\theta}=\frac{1}{J}\sum_{j\in\Upsilon_{i}}\theta_{j}$.
$\bar{G}_{j}=\frac{1}{n_{j}}\sum_{p=1}^{t}\left[y_{jp}^{*}\cdot\left(Q_{jp}-\mu^{d}\cdot Distance_{j}\right)\right]$,
$n_{j}=\sum_{p=1}^{t}y_{jp}^{*}$, $n^{d}=\sum_{j\in\Upsilon_{i}}\sum_{p=1}^{t}\left(y_{jp}^{*}\cdot Distance_{j}^{2}\right)$
and $G^{d}=\sum_{j\in\Upsilon_{i}}\sum_{p=1}^{t}\left[y_{jp}^{*}\cdot\left(Q_{jp}-\mu_{j}\right)\cdot Distance_{j}\right]$
. 

\subsection{Computation\label{subsec:Computation}}

The Gibbs sampling estimation of the simple (see $\mathsection$4.3.2)
and regression model (see $\mathsection$4.3.3) is carried out in
the same way: for each customer in each period, if new information
is available to the customer (i.e., the customer observes the service
quality of a shipment with AlphaShip), 1000 samples are iteratively
drawn from the full-conditional posterior distributions, where the
first 500 samples are discarded as burn-in. To speed up convergence,
for each period, we use the last period's estimate (e.g. $\mu_{j,t-1}^{E}$,
$\mu_{t-1}^{E}$, $\sigma_{p}^{E}$) as a starting point. Finally,
because customers learn service quality independently, parallel computing
(32 threads) is adopted, which significantly reduces the computational
time. Accordingly, each of the 32 threads generates the posterior
draws for a different customer. This code was implemented in Matlab,
and the longest running time was 50h on a 2.96-GHz Intel Xeon E5-2690
computer with 32 cores.

\section{Supplementary Material for Choice Model Estimation}

This section provides supplementary material for the computation details,
simulation experiments to test and validate our methodology and estimation
results of the shipping choice model. 

\subsection{Shipping Choice Model Estimation\label{subsec:Shipping-Choice-Model}}

The estimation of the demand arrival and shipping choice model parameters
(Equation (\ref{eq: Demand Arrival}) and (\ref{eq: RUM})) is implemented
using a simulated maximum likelihood estimation technique \citep{train_discrete_2009}.
The goal is to estimate $(i)$ the distribution of the intercept,
price and service quality sensitivity parameters, which are governed
by $\boldsymbol{\beta}$ and $\Omega$; and $(ii)$ the vector of
parameter, $\boldsymbol{\beta}^{X}$, related to the control variables.
The likelihood function is then given by: 
\[
L=\int\prod_{i}\prod_{t}\left[\left(\prod_{j\in\Upsilon_{i}}\left(\lambda_{i}m_{ij}P_{ijt}\right)^{y_{ijt}}\right)\left(1-\sum_{j\in\Upsilon_{i}}\lambda_{i}m_{ij}P_{ijt}\right)^{1-\sum_{j\in\Upsilon_{i}}y_{ijt}}\mid\boldsymbol{\theta}\right]dF\left(\boldsymbol{\theta}\right),
\]
where $\boldsymbol{\theta}$ is the vector of draws from a multivariate
normal distribution with mean $\boldsymbol{\beta}$ and $\boldsymbol{\Omega}$.
We resort to Halton draws to simulate the above integral over the
parameter spaces in order to keep the simulation error low \citep{train_discrete_2009}.
We run the simulated maximum likelihood estimation for 100 draws for
each of the individual-level parameters of the model.

\subsection{Simulation Experiment\label{subsec:Numerical-Experiment}}

In this subsection, we test the proposed methodology using simulated
data. We generate purchase and service experience data for 100 customers
and 50 periods. Specifically, each customer can ship from 2 to 6 routes
in the 50 periods, and we simulate his shipping demand in each period
from a multinomial distribution. Then, we simulate the service experience
(i.e., delays) data from a $N\left(0,1\right)$ distribution. In each
period, within the arrival of each new experience, the customer updates
his belief about delays following the simple Bayesian hierarchical
model and their prior beliefs are flat priors as we use in $\mathsection$\ref{Table: Pre-estimation Priors}.
Then, the beliefs about service quality (represented by $x_{1}$),
together with the intercept and another variable (represented by $x_{2}$)
are used in the utility function, where the values of $x_{2}$ are
generated from $N\left(0,1\right)$. We use random coefficients for
the intercept and transport delay $x_{1}$, and assume fixed (i.e.,
constant across customers) coefficients for $x_{2}$. Accordingly,
The coefficient of the intercept, $x_{1}$ and $x_{2}$ are generated
from a multivariate normal distribution with mean $\bar{\theta}=${[}-0.5,
0.3, -0.4{]} and diagonal variance matrix $\Sigma$ in which elements
equal to 0.6 and 0.5 for the intercept and $x_{1}$, respectively,
and the element equals to zero for $x_{2}$. 
\begin{figure}[tb]
\begin{centering}
\includegraphics[width=0.6\columnwidth]{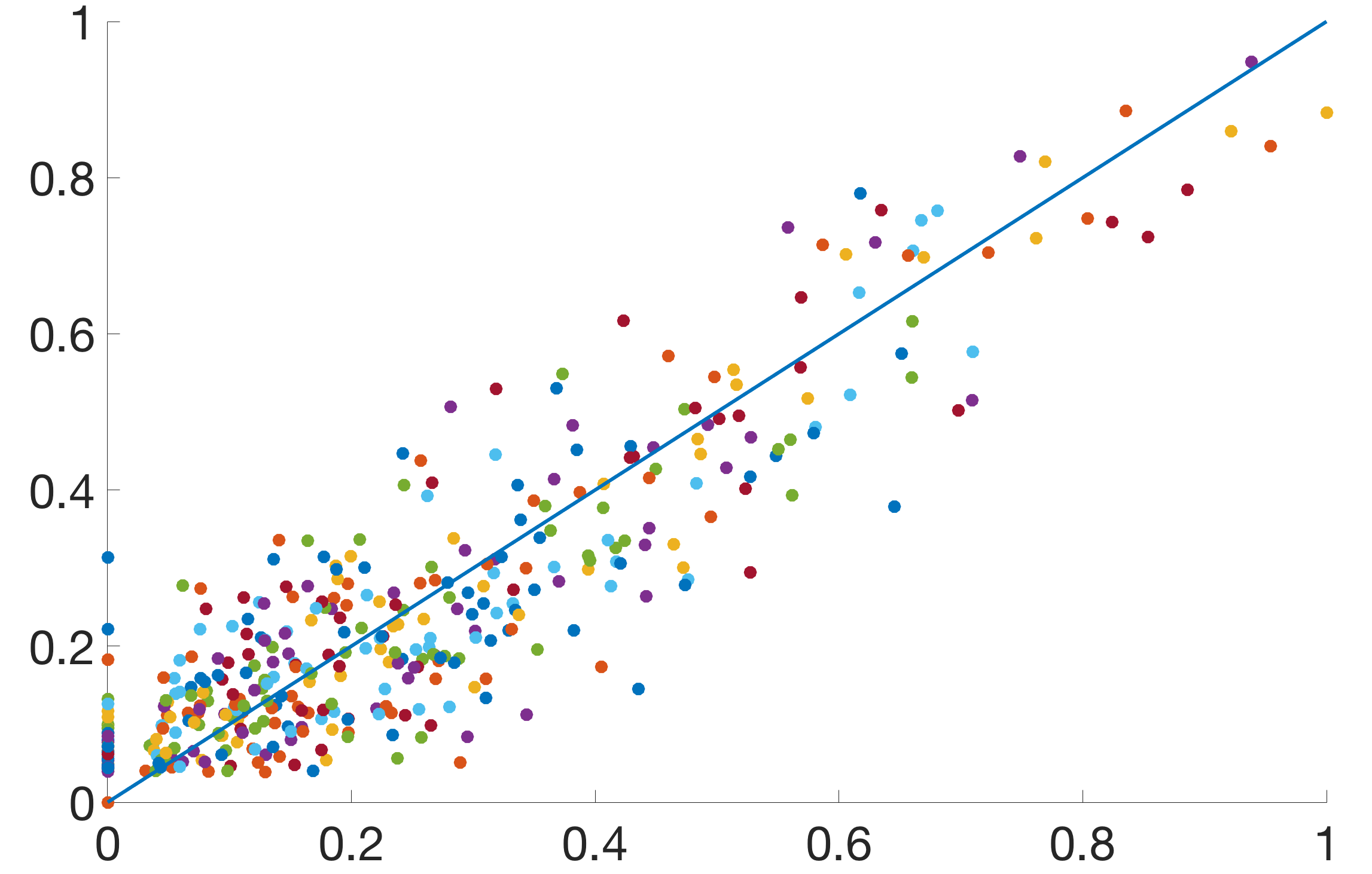}
\par\end{centering}
\caption{Actual (Y axis) vs. Estimated (X axis) Values of the Multinomial Distribution
Parameters}
\label{Figure: simulation estimate multinomial}
\end{figure}
\begin{table}[tb]
\centering{}\caption{Actual vs. Estimated Values of the Discrete Choice Model Parameters}
\label{Table: simulation estimate discrete choice model}%
\begin{tabular}{>{\raggedright}p{1.5cm}>{\centering}p{2.5cm}>{\centering}p{1.5cm}>{\raggedleft}p{1.5cm}}
\hline 
 & \textsf{\small{}True value} & \multicolumn{2}{c}{\textsf{\small{}Estimated value}}\tabularnewline
 &  & \textsf{\small{}Mean} & \textsf{\small{}Std. err.}\tabularnewline
\hline 
\textsf{\small{}$\theta_{1}$} & \textsf{\small{}-5e-1} & \textsf{\small{}-5.5e-1} & \textsf{\small{}3.3e-2}\tabularnewline
\textsf{\small{}$\theta_{2}$} & \textsf{\small{}3e-1} & \textsf{\small{}2.8e-1} & \textsf{\small{}4.8e-2}\tabularnewline
\textsf{\small{}$\theta_{3}$} & \textsf{\small{}-4e-1} & \textsf{\small{}-4.2e-1} & \textsf{\small{}3.9e-2}\tabularnewline
\textsf{\small{}$\Omega\left(\theta_{1}\right)$} & \textsf{\small{}6e-1} & \textsf{\small{}5.7e-1} & \textsf{\small{}4.2e-2}\tabularnewline
\textsf{\small{}$\Omega\left(\theta_{2}\right)$} & \textsf{\small{}5e-1} & \textsf{\small{}4.6e-1} & \textsf{\small{}4.3e-2}\tabularnewline
\hline 
\end{tabular}
\end{table}

Because there are too many parameters in the multinomial distribution
demand model (see $\mathsection$\ref{subsec:Demand-Arrival}), instead
of listing all the estimation values in a table, we plot the estimates
versus the true values into Figure \ref{Figure: simulation estimate multinomial}.
As we can see from Figure \ref{Figure: simulation estimate multinomial},
the estimates are close to the true value (i.e., the dots spread evenly
around the $y=x$ line). Table \ref{Table: simulation estimate discrete choice model}
provides the estimates of the discrete choice model parameters, where
we can see the estimates are close to the true values and the estimation
standard errors are very small compared to the true value. The results
in Figure \ref{Figure: simulation estimate multinomial} and Table
\ref{Table: simulation estimate discrete choice model} show that
our method recovers well the original parameters. 

\subsection{Supplementary Material for Results}

In Table \ref{Table: Model Fitting LL} are the log-likelihoods under
different combinations of the learning model and the quality function
specification. 
\begin{table}[tb]
\caption{Log-Likelihood of Alternative Model Specifications}
\label{Table: Model Fitting LL}
\centering{}%
\begin{tabular}{>{\raggedright}p{3.1cm}|>{\centering}p{2cm}>{\centering}p{2cm}>{\centering}p{2cm}>{\raggedleft}p{2cm}}
\hline 
 & \textsf{\footnotesize{}Benchmark Model 1 (short memory)} & \textsf{\footnotesize{}Benchmark Model 3A (information pooling)} & \textsf{\footnotesize{}Benchmark Model 2 (independent learning)} & \textsf{\footnotesize{}Simple Hierarchical Model}\tabularnewline
\hline 
\textsf{\footnotesize{}S{*}} & \textsf{\footnotesize{}-53647.9} & \textsf{\footnotesize{}-53,628.4} & \textsf{\footnotesize{}-53,626.4} & \textsf{\footnotesize{}-53,623.6}\tabularnewline
\textsf{\footnotesize{}A} & \textsf{\footnotesize{}\textemdash \textendash{}} & \textsf{\footnotesize{}-53,613.6} & \textsf{\footnotesize{}-53,593.5} & \textsf{\footnotesize{}-53,582.1}\tabularnewline
\textsf{\footnotesize{}A + ERA} & \textsf{\footnotesize{}\textemdash \textendash{}} & \textsf{\footnotesize{}-53,588.0} & \textsf{\footnotesize{}-53,572.1} & \textsf{\footnotesize{}-53,569.7}\tabularnewline
\textsf{\footnotesize{}A + ERA + BUA} & \textsf{\footnotesize{}\textemdash \textendash{}} & \textsf{\footnotesize{}-53,563.3} & \textsf{\footnotesize{}-53,557.7} & \textsf{\footnotesize{}-53,544.2}\tabularnewline
\textsf{\footnotesize{}A + ERA + BUA + Q} & \textsf{\footnotesize{}\textemdash \textendash{}} & \textsf{\footnotesize{}-53,563.8} & \textsf{\footnotesize{}-53,557.6} & \textsf{\footnotesize{}-53,544.0}\tabularnewline
\hline 
\multicolumn{5}{>{\raggedright}p{11.1cm}}{\textsf{\scriptsize{}{*}We use S for symmetric, A for asymmetric,
ERA for experience risk averse, BUA for belief uncertainty averse,
Q for quadratic. }}\tabularnewline
\end{tabular}
\end{table}

\section{Supplementary Material for Policy Simulations\label{sec:Supplementary-Application} }

In addition to the policy simulation study provided in $\mathsection$\ref{sec: Applications},
we also consider the effect of an increase in experience variability
(i.e., more volatile service experiences) on a certain route. 

Same as in $\mathsection$\ref{sec: Applications}, we consider customers
with frequent routes $A$ and $B$ and whose demand on the two routes
follows a multinomial $\left(0.5,0.5\right)$ distribution. Transport
delays on these two route, $Q_{A}$ and $Q_{B}$ respectively, follow
the same normal distribution, $Q_{A}\sim N\left(0,5^{2}\right)$ and
$Q_{B}\sim N\left(0,5^{2}\right)$. Customers decide whether to ship
based on the utility of shipping compared to that of the outside option
and then update their quality beliefs if new information is acquired
(i.e., if they used AlphaShip services).\footnote{When calculating utilities we set other exogenous variables (e.g.,
price) to their empirical means.} 
\begin{figure}[tb]
\begin{centering}
\includegraphics[width=0.6\columnwidth]{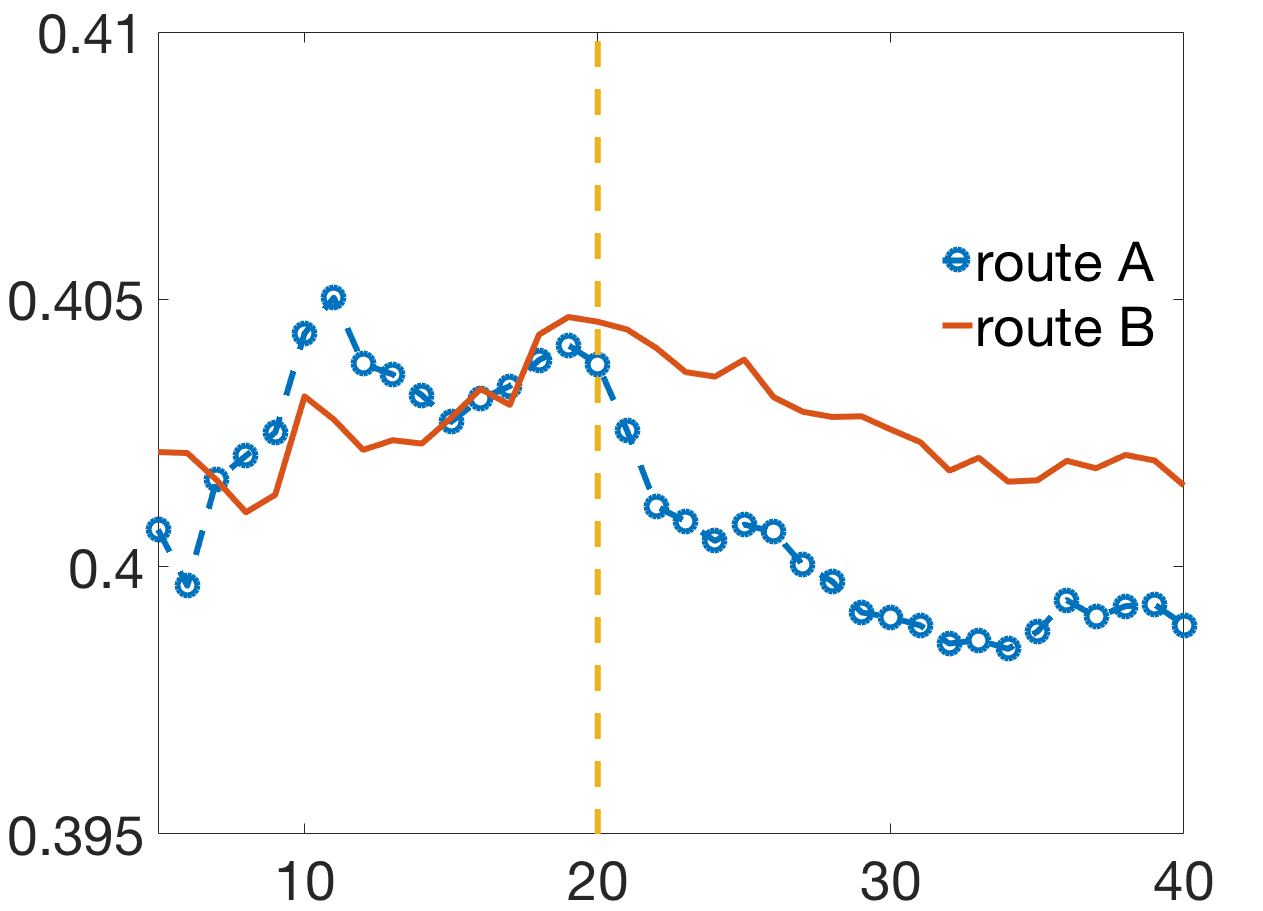}
\par\end{centering}
\centering{}\caption{Purchase probability changes on route A and B under variance deterioration. }
\label{Figure: Appendix-Counterfactual}
\end{figure}

Figure \ref{Figure: Appendix-Counterfactual} depicts the average
purchase probabilities for routes $A$ and $B$. For this scenario,
we consider an increase in the service quality variability for route
$A$ which takes place in period 20: $Q_{A}\sim N\left(0,10^{2}\right)$.
This change makes customers anticipate more variability in service
quality not only for that route for also for route $B$. This greater
variability in turn decreases the purchase probability on route $B$.
Even though both route $A$ and $B$ share the same experience variability
and belief uncertainty, the additional drop of average purchase probability
on route $A$ is caused by the asymmetric effect of delays and earliness.
Specifically, because service quality variability on $A$ becomes
larger after period 20, both the magnitude of delays and early deliveries
experienced on route $A$ increase. Although the service quality distribution
is symmetric, customers are more sensitive to delays rather than earliness.
Hence, as the magnitude of delays increases, the average purchase
probability on route $A$ decreases more than that on route $B$.
In terms of revenues per customer, in period 40, the direct loss from
the increased transport delay variance of route $A$ is \$18.3 on
route $A$ and \$8.0 on route $B$. This implies that indirect losses
account for 31\% of the total revenue loss. 

\end{APPENDICES} 
\end{document}